\documentclass[aps,prd,twocolumn,amsmath,amssymb,floatfix,nofootinbib]{revtex4}

\usepackage{graphicx}
\usepackage{dcolumn}
\usepackage{bm}
\usepackage{color}

\newcommand{\be}{\begin{eqnarray}}

\newcommand{\ee}{\end{eqnarray}}

\usepackage{subfig}
\def\<{\langle}
\def\>{\rangle}

\begin{document}

\title{Joint resonant CMB power spectrum and bispectrum estimation}
\author{P.~Daniel Meerburg$^{1}$}
\author{Moritz M\"{u}nchmeyer$^{2,3}$}
\author{Benjamin Wandelt$^{2,3,4,5}$}
\affiliation{$^1$CITA, University of Toronto, 60 St. George Street, Toronto, Canada}
\affiliation{$^2$Sorbonne Universit\'{e}s, UPMC Univ Paris 06, UMR7095}
\affiliation{$^3$CNRS, UMR7095, Institut d'Astrophysique de Paris, F-75014, Paris, France}
\affiliation{$^4$Lagrange Institute (ILP) 98 bis, boulevard Arago 75014 Paris France}
\affiliation{$^5$Departments of Physics and Astronomy, University of Illinois at Urbana-Champaign, Urbana, IL 61801, USA}

\begin{abstract}
We develop the tools necessary to assess the statistical significance of resonant features in the CMB correlation functions, combining power spectrum and bispectrum measurements. This significance is typically addressed by running a large number of simulations to derive the probability density function (PDF) of the feature-amplitude in the Gaussian case. Although these simulations are tractable for the power spectrum, for the bispectrum they require significant computational resources. We show that, by assuming that the PDF is given by a multi-variate Gaussian where the covariance is determined by the Fisher matrix of the sine and cosine terms, we can efficiently produce spectra that are statistically close to those derived from full simulations. By drawing a large number of spectra from this PDF, both for the power spectrum and the bispectrum, we can quickly determine the statistical significance of candidate signatures in the CMB, considering both single frequency and multi-frequency estimators. We show that for resonance models, cosmology and foreground parameters have little influence on the estimated amplitude, which allows to simplify the analysis considerably. A more precise likelihood treatment can then be applied to candidate signatures only. We also discuss a modal expansion approach for the power spectrum, aimed at quickly scanning through large families of oscillating models. 
\end{abstract}

\maketitle

\tableofcontents

\section{Introduction}

We have come a long way in our endeavor to understand the early Universe. Modern cosmology allows us to probe the history of our Universe all the way up to GUT scales, when the Universe was only a fraction of second old \cite{WMAPfinal2012,PlanckInflation2013,PlanckNGs2013,PlanckNGs2015,PlanckInflation2014,PlanckCosmoPars2015,ACT2014}. The challenge we face is to build a consistent theory of quantum gravity and propose reliable tests. Cosmology provides us with observational measures that have the potential to test models of quantum gravity. However, probing the early Universe has its limitations; damping of acoustic waves below $k\sim 0.1$ Mpc$^{-1}$ in the Cosmic Microwave background and the non-linear scale today of order $k_{\rm NL}\sim 0.1$ Mpc$^{-1}$ prevent us from reconstructing a full picture. Future probes and a better understanding of non-linear physics \cite{LSSnonGaussianity2014} will hopefully allow us to extend our view into the past, but even now we should try and utilize all accessible information in order to build a comprehensive picture of the early Universe and constrain models of quantum gravity. One way forward is to combine multiple $n$-point statistics to determine the joint likelihood of given a model. 

In this paper we build a framework for combining power spectrum and bispectrum observables to obtain a joint constraint on resonant features in primordial correlation functions. These features are predicted by various models, but as an example we will focus on axion monodromy \cite{NaturalInflation1990,MonodromySilverstein2008,MonodromyFlauger2009,ResNGsFlaugerPajer,DriftingOscillations2014,LargeBispectrumGreen2012}. Another interesting model that predicts a very similar CMB power spectrum shape is unwinding inflation \cite{UnwindingInflation2013}. It is straightforward to extend the framework presented in this paper to other models that predict correlated oscillating features in primordial n-point correlation functions \cite{NGFeaturesChen2007,StepTheoryJoyEtAl2007,EOSchangeBattefeldEtAl2010,StepFeatureAdshead,InitialStateOriginalHolman2007,CorrelatedFeaturesCs,NonBDBispectrum2010,NonBDBispectrum2010b,FeaturesFromHeavyPhysicsAna2011,NonBDBispectrum2009,FeaturesFromSoundSpeedAna2014,2fieldFeatures2013,ResonantAndNonBDChen2010}. The main difficulty in estimating the joint significance is that one has to sample the joint probability distribution function (PDF) of the estimators. If the dependence on cosmological parameters and foreground parameters is taken into account, this cannot be done analytically. Alternatively, a large number of simulations of the Gaussian Universe can be performed and the PDF can be reconstructed by applying the power- and bispectrum estimators to these simulations. Unfortunately, simulating a Gaussian map and applying the bispectrum estimator to it, is computationally expensive; each simulation takes $\mathcal{O}(10^3)$ to $\mathcal{O}(10^4)$ CPU hours (depending on resolution parameters), rendering this route computationally challenging. Instead, we propose a much faster way to derive the PDF, by using the correlation matrix of the estimator that can be computed analytically. We assume that the PDF is approximately described by a multivariate Gaussian in the the spectrum of amplitudes, as we will explain in more detail below. Once the correlation matrix has been calculated for both the power spectrum and bispectrum, one can quickly draw simulated spectra from the PDF. Based on this approximation, we can sample PDFs for different estimators and obtain the related look-elsewhere effect, without having to rely on a massive amount of Monte Carlo simulations. We argue that our approximation, while neglecting some systematics, is good enough for its primary purpose, which is to assess significances in combined estimates from the data. The PDFs necessary to calculate combined significances have been previously studied nicely in Ref.~\cite{ShellardBispectrumPsJoined2014,Fergusson:2014tza}, which focussed on linear feature model oscillations, and the statistics proposed in these papers have been applied to the Planck data in~\cite{PlanckNGs2015} for both linear and logarithmic oscillations. The main new contribution in this paper is a way to sample these PDFs as well as a study of the influence of foreground and cosmological parameters on the likelihood in the case of resonance models. We also propose a modal expansion of the power spectrum (similar to our bispectrum approach), which allows to quickly scan a large range of oscillating models. This is in particular useful when further parameters (like the frequency drifting in Ref.~\cite{DriftingOscillations2014}) are taken into account.

This paper is organized as follows. In Sec.~\ref{sec:models} we review the predictions of the class of models that we are primarily interested in. In Sec.~\ref{sec:estimators} we describe our estimators and describe our approximation of the estimator PDF. The core results of this paper are given in Sec.~\ref{sec:combined}, where we evaluate the look-elsewhere effect for different combined estimators.

\section{Review of resonance models}
\label{sec:models}

Many models of inflation share the property of oscillations in the power spectrum and bispectrum of the primordial density perturbations. In this paper we are concerned with models of inflation that predict logarithmic oscillations in the wave number $k$, the so called resonance models \cite{ResonantAndNonBDChen2010}. The most well-studied UV-complete realization of this model is axion monodromy inflation \cite{MonodromySilverstein2008,MonodromyFlauger2009, BICEPoscillations2014}. However, more generally, a broken discrete shift symmetry \cite{EFTOscillations2011} leads to the same $k$-dependence. Below we review these shapes, and set up the notation for the following sections.

For both axion monodromy (AXI) and unwinding inflation (UI) the default form of the power spectrum can be captured with the following template shape
\be
P(k)  = P^0 \left(1+A_P \cos \left(\omega \log k/k_* +\phi_P\right)\right).
\label{eq:powerspectra1}
\ee
AXI is a UV complete model which features a shift symmetry (thus protecting the slow-roll potential from large quantum corrections), and which allows for super Planckian field displacements and thus for potentially observable levels of gravitational waves \cite{BICEPoscillations2014}. Details of the compactification setup influence the predictions, however the shape of the power spectrum is fairly generic. Generally the frequency $\omega$ is related to the axion decay constant $f$  and the tensor-to-scalar amplitude $r_*$ as $\omega \propto \sqrt{r_*}/f$. The amplitude is also proportional to the axion decay constant, but depends on further parameters. The predicted range of frequencies for this shape is $0<\omega<\mathcal{O}(10^2)$ in an EFT setting \cite{EFTOscillations2011} but could be higher in a UV completion like axion monodromy. We will consider frequencies up to $\omega=1000$.

UI is based on the premise of an eternally inflating metastable false vacuum which transitions via charged brane bubble formation to a flux discharge cascade. This cascade mimics slow-roll inflation. The vacuum energy steadily ÒunwindsÓ over time. Inflation ends and reheating occurs with the self-annihilation of the brane into radiation once most or all vacuum energy is discharged. Log-spaced oscillations are naturally produced when the flux associated with the inflaton scalar unwinds on cycles in compact directions. Again, this leads to an effective oscillatory potential. The amplitude of the correction to the usual power spectrum depends on the number of cycles and details of the model. 

For AXI the lowest order bispectrum was computed in Ref. \cite{ResNGsFlaugerPajer} and is given by
\be
\label{eq_oscispectrum1} 
B_\Phi(k_1,k_2,k_3) = \frac{P_0^2 A_B}{(k_1k_2k_3)^2} \sin\left({\omega \log \frac{k_t}{k_*} + \phi_B}\right).
\ee
Both the power spectrum and the bispectrum are predicted to have the same frequency $\omega$, and in this paper the pivot scale is set to $k_*=1~\mathrm{Mpc}^{-1}$. No analytic computation exists for the bispectrum of the unwinding model. The leading mechanism that produces non-Gaussianity is the reduced speed of sound, which generally leads to non-Gaussianities of the equilateral type, so one might expect a combination of the equilateral template and resonant features. As the precise template shape has not yet been derived, we will use the resonant bispectrum as our default shape in the rest of this paper. We expect however that all the presented techniques will also be applicable in a straightforward way to unwinding inflation.


For an optimal combined measurement, it is necessary to know the predicted relation of power spectrum and bispectrum amplitudes $r=A_P/A_B$ as well as the predicted relation of phases $\Delta \phi=\phi_P-\phi_B$. For example, if one would find a large amplitude at the same frequency in power spectrum and bispectrum, a correspondence between phases would give additional significance not included in the amplitudes. If on the other hand the model does not predict $r$ and $\Delta \phi$, one has to scan over these parameters which leads to an additional look-elsewhere effect. We will discriminate these cases, as was previously done in Ref.~\cite{ShellardBispectrumPsJoined2014,Fergusson:2014tza}.

One can generalize the models above to include multiple frequencies, which could be generated by multiple axions or multiple instantons in axion monodromy. We define our multi-frequency shape for $M$ frequencies as 
\be
P(k)  = P^0 \left(1+ \sum_i^M A_{P,i} \cos \left(\omega_i \log k/k_* +\phi_{P,i}\right)\right),
\label{eq:powerspectra2}
\ee
and similarly for the bispectrum
\be
\label{eq_oscispectrum2} 
B_\Phi(k_1,k_2,k_3)& = & \frac{P_0^2}{(k_1k_2k_3)^2} \times \nonumber \\
&& \sum_i^M A_{B,i} \sin\left({\omega_i \log \frac{k_t}{k_*} + \phi_{B,i}}\right),
\ee
with amplitude ratios $r_i$ and phase differences $\Delta \phi_i$. 

It was realized in Ref.~\cite{DriftingOscillations2014} that the oscillation frequency in axion monodomy can drift, which leads to a modified power spectrum shape. Two parameterizations of drifting oscillations have been proposed in Ref.~\cite{DriftingOscillations2014} and constrained with the Planck data \cite{PlanckInflation2014}. The ``semi-analytic'' template is given by
\begin{equation}\label{eq:temp1}
P(k)  = P^0 \left(1+A_P \cos\left[\omega \left( \ln(k/k_\star) \right)^{\frac{1}{2}(p_f+1)}+ \phi_P \right]\right)\,,
\end{equation}
which reduces to the simplest shape for $p_f=1$. The additional parameter $p_f$ is expected to be a rational number of modest size, and was set to prior values $-0.75<p_f<1$ in the Planck paper.   The ``analytic template'' results from an expansion of the semi analytic template to third order in the logarithm:
\be \label{eq:lnn}
P(k)  &=& P^0 A_P \cos\left(\omega\left[\ln(k/k_\star)+\sum\limits_{n=1}^2 c_n \ln^{n+1}(k/k_\star)\right] + \right.  \nonumber \\
&&   \phi_P \Bigg) + P_0. 
\ee
This form has one additional free parameter ($c_1$ and $c_2$ replace $p_f$) and can therefore represent more general shapes than the previous one. We will briefly comment on how to determine the associated look-elsewhere effect when adding these free parameters. So far no corresponding bispectrum shapes with drifting oscillations have been derived in the literature.


\section{Estimators for power spectrum and bispectrum}
\label{sec:estimators}

Power spectrum and bispectrum parameter estimation is usually done in a different setting, and there have been several methods proposed to obtain reliable high frequency oscillation estimators. In this section we describe our estimators for both power spectrum and bispectrum, and make an analytic approximation to the estimator probability density function to be used in the following sections.

\subsection{Power spectrum likelihood estimator}
\label{sec:estimators_p}

The search for oscillations in the power spectrum is usually done by extending the CMB likelihood exploration with additional parameters describing the oscillating shape, and modifying your favorite Boltzmann code (e.g. CAMB \cite{cosmomc} or CLASS \cite{CLASS}) to calculate the resulting oscillating power spectra. There are several methods and codes available to perform this search for rapid features in the power spectrum. The two main obstacles in constraining features are 1) the computational cost of computing a high resolution spectrum for each model parameter and 2) the irregular likelihood and corresponding convergence problems. We will use the method developed in Ref.~\cite{Meerburg2014a,Meerburg2014b}. In order to solve the first obstacle, the authors proposed to separate the rapid oscillating part from the smoothly varying part, i.e. 
\be
\mathcal{C}_{\ell}^{\rm tot} = \mathcal{C}_{\ell} + \mathcal{C}_{\ell}^{\rm osc}.
\label{eq:ClPert}
\ee
This separation is justified because generally $\mathcal{C}_{\ell}^{\rm osc}/\mathcal{C}_{\ell} \ll 1$. The oscillating part of the spectrum can be precomputed and stored for interpolation using a modified version of CAMB which increases sampling in $k$ and computes all $\ell$. In principle this approach can be improved by varying the oscillating part with respect to the 'slow' cosmological parameters in order to include degeneracy effects. Again, these derivatives can be precomputed. This code uses  the Multinest sampler \cite{Feroz2009,Feroz2013} to deal with the irregular likelihood. Because of the pre-computation of the oscillating contributions, this code is very fast; varying all cosmological parameters leads to sufficient convergence within a large prior range in frequency space in approximately $400$ CPU hours \cite{MeerburgOscillations2014}. The code can be iteratively improved as explained by taking higher order derivatives of the $\mathcal{C}_{\ell}^{\rm osc}$ with relatively little additional computational cost. It also works for any type of oscillating spectrum; both linear, logarithmically and local features can similarly be treated; all that has to be changed is the precomputed files, which require very little time to compute. 

A more elaborate and exhaustive search was performed in Ref.~\cite{PlanckInflation2014}. All cosmological parameters are varied using the Polychord sampler \cite{Polychord} in combination with a  high resolution version of CAMB. We experimented with some of the settings that allow to accurately compute spectra with $\omega = \mathcal{O}(10^2)$, and found it slows down a single power spectrum computation by a factor 3-4. For very high frequencies (considered in \cite{PlanckInflation2014}), at low $\ell$, the sampling in $k$ needs to be increased to avoid glitches \cite{PSOscillations2011}, while at high $\ell$ pre-set sampling in $\ell$ is insufficient and one reaches a point where {\it all} $\ell$ need to be computed in order to resolve the oscillations. For these high frequencies, we found that a single power spectrum takes as much as 10 times as long. We were not able to build a working code; we found slow convergence to be the major culprit. 


\subsection{Influence of cosmology and foreground parameters}

The power spectrum and bispectrum analysis differs in a crucial manner, which is that the bispectrum is traditionally done for a fixed cosmology and foreground model, while in the power spectrum case one explores the full likelihood. The main reason for this difference is that exploring a bispectrum likelihood is computationally prohibitive. For a fair comparison of bispectrum and power spectrum, one may therefore want to fix the cosmology and foregrounds also in the power spectrum. In this case the analysis simplifies considerably. Here we study how fixing these parameters influences the power spectrum results. 

We first make some intuitive arguments as to what extent fixing the cosmological parameters influences the result. In either power-spectrum or bispectrum, neither cosmology nor foreground parameters are expected to introduce high frequency oscillations in the power spectrum. Therefore for rapid oscillations, fixing these parameters should not be problematic. On the other hand, for oscillations near the BAO frequency, fixing the cosmology could bias the result. For the logarithmic oscillations that we consider here, the frequency $\omega$ of the shape varies over the multipole range so that BAOs in principle contribute to many oscillating shape frequencies. On the other hand, cosmological parameters are constrained by a multitude of observations and are known to good precision. For linear oscillations, the influence of fixing the cosmological parameters on the estimated oscillation amplitude in the power spectrum has been studied in Ref.~\cite{ShellardBispectrumPsJoined2014} by use of simulated maps, finding that indeed BAO frequencies are biased while higher frequency amplitude estimates are cosmology independent. Ref. ~\cite{Fergusson:2014tza} showed that this conclusion also holds for the Planck data, again for linear oscillations. For the bispectrum such a study would be computationally challenging. One may expect that because the bispectrum sums over many multipole combinations (the $k$-triangles), the effect of BAO frequencies should be less pronounced than in the power spectrum. Finally, as long as no significant amplitude is observed in the data, one can have the practical point of view that a significant BAO bias is not observed. 

\begin{figure*}
\resizebox{0.9\hsize}{!}{
\includegraphics{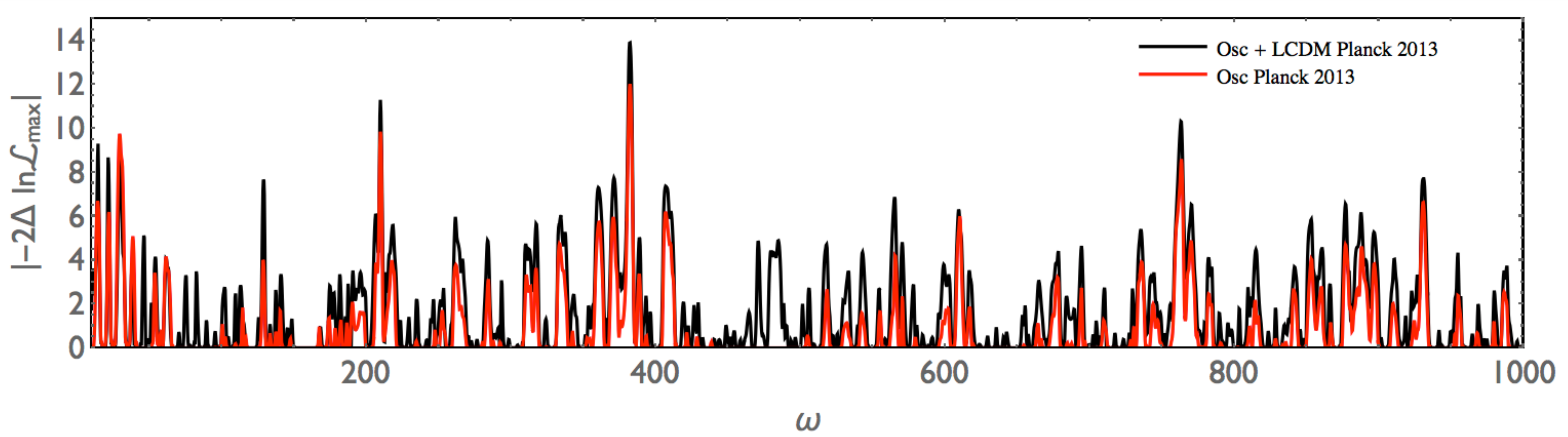}
}
\resizebox{0.9\hsize}{!}{
\includegraphics{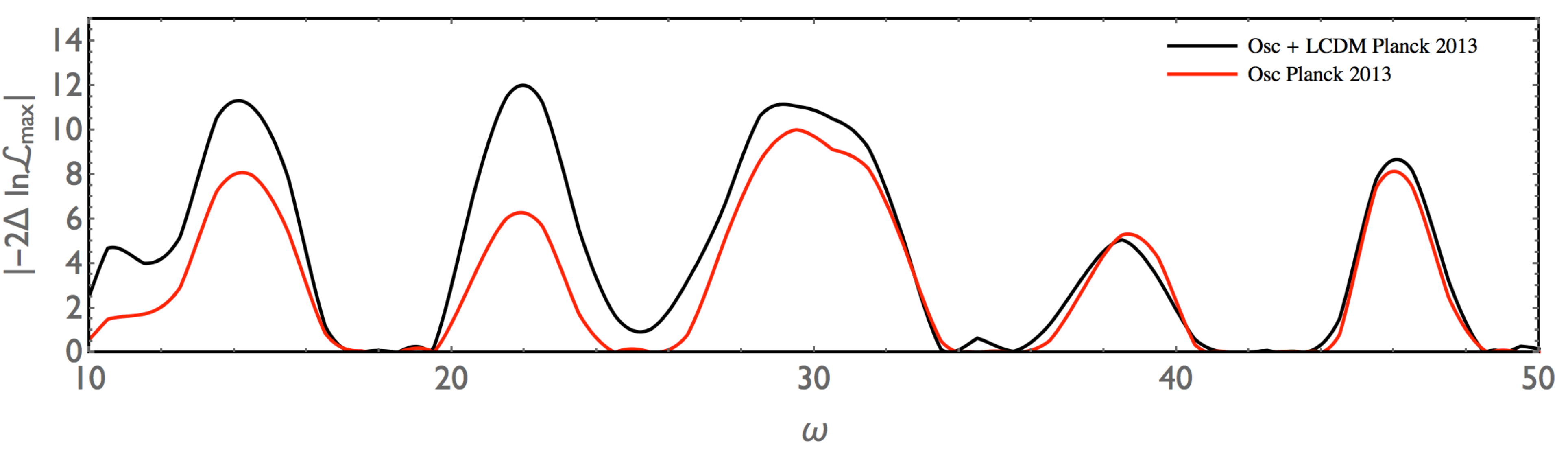}
}
\caption{Top: Power spectrum analysis using the template shape of Eq.~\eqref{eq:powerspectra1} (black) using the approximation of Eq.~\eqref{eq:ClPert} and the Planck 2013 temperature likelihood. We compare this to a similar analysis where we fix all $\Lambda$CDM parameters (red). It is clear that over the range of scanned frequency there are no strong corrections. The overall improvement is reduced, but all the important features are recovered. Note that the improvement is systematically lower, except at the low frequencies, indicating that cosmological parameters are degenerate with some of the effects for those frequencies.  Bottom: The same analysis with a tighter prior on the frequency. Even on low frequencies the difference is negligible justifying fixing all parameters for log oscillating signals. }
\label{fig:MultinestCompare}
\end{figure*}

Let us now examine this question in an quantitative way for the power spectrum. As an example, we show results of the 2013 Plank temperature data \cite{Planck2013Likelihood} for the simplest template shape in the top of Fig.~\ref{fig:MultinestCompare}. We use default priors \cite{PlanckInflation2014} on the cosmological parameters, while $1\leq \omega \leq 1000$, $ 0 \leq A_p \leq 0.8$ \footnote{Note that observationally $\mathcal{C}_{\ell}^{\rm osc}/\mathcal{C}_{\ell} \ll 1$ even though the primordial amplitude can be as large as 0.8. The largest amplitudes are only possible for very high frequencies, which significantly suppresses the total power after projection. Very similar to the bispectrum \cite{OptimalEstimator2014}, the projected amplitude of the power spectrum is suppressed (see Appendix) as $A_{\rm CMB} \sim A_P \omega^{-1}$; for very high frequencies and small scales it is possible to still have a very large primordial amplitude. Our expansion is based on the assumption that $\mathcal{C}_{\ell}^{\rm osc}/\mathcal{C}_{\ell} \ll 1$ and holds over the scanned frequency range. } and $-\pi \leq \phi \leq \pi$. We precomputed the $ \mathcal{C}_{\ell}^{\rm osc}$ with $\Delta \omega =0.5$ and the code interpolates for frequencies in between. We used only temperature data. For more advanced shapes the precomputed files need to include variations in e.g. $p_f$ and $c_1$ and $c_2$. 

We show the same analysis with fixed $\Lambda$CDM parameters in same Fig.~\ref{fig:MultinestCompare}. We find negligible difference indicating that log oscillations have very little correlation with the other cosmological parameters\footnote{There is a dependence of the estimated frequency and phase that depends on the cosmological parameters; this shift the actual frequency. It would be wise to vary all parameters in the vicinity of a possible detection to establish the correct frequency and phase \cite{Meerburg2014a}. }. We also scanned over a tighter prior to determine if there were indications of correlations on large (BAO) scales, as was shown to be the case for linear oscillations \cite{ShellardBispectrumPsJoined2014}. The results are shown at the bottom of Fig.~\ref{fig:MultinestCompare}. Note that the semi-full analysis takes over 400 CPU hours, while for a fixed $\Lambda$CDM cosmology convergence is achieved within 1 CPU hour.

\subsection{Power spectrum modal estimator}
\label{sec:estimators_modal}

Here we present an estimator for oscillations in the power spectrum based on a Fourier decomposition in linear oscillations. Our approach has several advantages with respect to other methods. First, scanning over many model parameters (e.g. frequency drifting) becomes computationally easier, since one only has to estimate the different modes once. Second, it is possible to remove specific linear modes that may induce systematics, for example at BAO frequencies. Third, it is easy to put power spectrum and bispectrum analysis on the same footing. If a significant peak were found, one would then use a full likelihood as in Sec.~\ref{sec:estimators_p} in the vicinity of that peak. Unlike a full likelihood exploration, we will have to fix cosmology and foreground parameters. 


We split the primordial power spectrum into the power law and the oscillating part as $ P(k) =  P_0(k) + p_{\rm NL} \delta P(k)$. Any power spectrum of the form $\delta P(k)$ can be developed in a Fourier series as
\be
\delta P(k) = \sum_{n=0}^N \left(  a_n \cos{\frac{2 \pi n k}{\Delta k}} + b_n \sin{\frac{2 \pi n k}{\Delta k}} \right),
\ee
where $\Delta k$ is the supported interval and the real Fourier coefficients are
\be
a_n &=& \frac{2}{\Delta k} \int_{k_{\rm min}}^{k_{\rm max}} dk \; \delta P(k) \cos{\frac{2 \pi n k}{\Delta k}} \\
b_n &=& \frac{2}{\Delta k} \int_{k_{\rm min}}^{k_{\rm max}} dk \; \delta P(k) \sin{\frac{2 \pi n k}{\Delta k}}.
\ee
\begin{figure*}
\resizebox{0.9\hsize}{!}{
\includegraphics{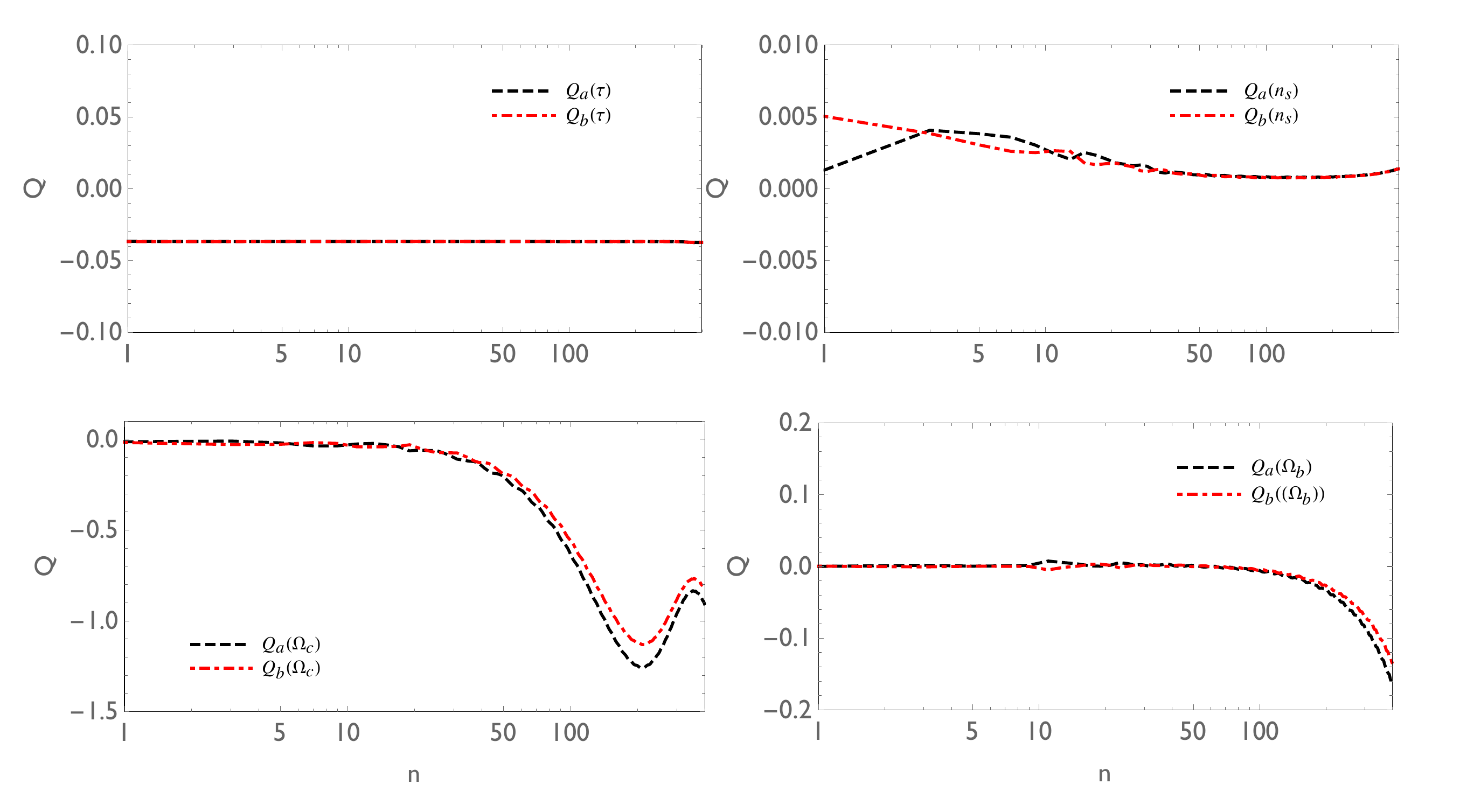}
}
\caption{Induced error in estimated modes $\hat{a}_n$ and $\hat{b}_n$ by fixing $\tau$, $n_s$, $\Omega_c$ and $\Omega_b$. }
\label{fig:errormodal}
\end{figure*}

For a fixed cosmology (i.e. using a fixed theoretical power spectrum $C^0_l$), the estimators for the modal coefficients are then given by
\be
\label{modcoeff}
\hat{a}_n&=&\frac{1}{\mathcal{N}_a}\delta C^{n,\cos}_{\ell_1}\Delta^{-1}_{\ell_1\ell_2}\left(\hat{C}_{\ell_2}-C^0_{\ell_2}\right),\\
\hat{b}_n&=&\frac{1}{\mathcal{N}_b}\delta C^{n,\sin}_{\ell_1}\Delta^{-1}_{\ell_1\ell_2}\left(\hat{C}_{\ell_2}-C^0_{\ell_2}\right)
\ee
such that 
\be
a_n  &=& \frac{a_n}{\mathcal{N}_a}\delta C^{n,\cos}_{\ell_1}\Delta^{-1}_{\ell_1\ell_2}\delta C^{n,\cos}_{\ell_2}, \\
b_n&=& \frac{b_n}{\mathcal{N}_b}\delta C^{n,\sin}_{\ell_1}\Delta^{-1}_{\ell_1\ell_2}\delta C^{n,\sin}_{\ell_2},
\ee
with 
\be
\mathcal{N}_a &=& \delta C^{n,\cos}_{\ell_1}\Delta^{-1}_{\ell_1\ell_2}\delta C^{n,\cos}_{\ell_2},\\
\mathcal{N}_b & =&\delta C^{n,\sin}_{\ell_1}\Delta^{-1}_{\ell_1\ell_2}\delta C^{n,\sin}_{\ell_2},
\ee
similarly to the reconstruction of $f_{\rm NL}$. 
Here $\delta C^n_{l_1}$ is the CMB power spectrum generated by a mode $\cos{\frac{2 \pi n k}{\Delta k}}$ or $\sin{\frac{2 \pi n k}{\Delta k}}$. The fiducial CMB power spectrum $C^0_{\ell}$ is the one generated by the power law primordial power spectrum $P_0(k)$. The estimator for the parameter $p_{\rm NL}$ is then
\be
\hat{p}_{\rm NL} = \frac{1}{\mathcal{N}_{\rm tot}} \left(\mathcal{N}_a  a_n \hat{a}_n +  \mathcal{N}_b b_n \hat{b}_n \right)
\ee
with norm $\mathcal{N}_{\rm tot} = F_{PP}$ and Fisher matrix 
\be
F_{PP'} &=&  \left(\sum_n a_n \delta C^{n,\cos}_{\ell_1} + b_n \delta C^{n,\sin}_{\ell_1} \right)\nonumber \\
&& \Delta^{-1}_{\ell_1\ell_2}   \left(\sum_m a'_m \delta C^{m,\cos}_{\ell_2} + b'_m \delta C^{m,\sin}_{\ell_2} \right)
\ee
where $\Delta^{-1}_{\ell_1\ell_2}$ is the covariance matrix of the experiment. This can be written in matrix form schematically as
\be
F_{PP'} = \mathbf{a_n F^{mode}_{nm} a_m}.
\ee
which allows fast evaluation of the Fisher matrix after the mode Fisher matrix is evaluated. An amplitude $p_{\rm NL}$ can be estimated with variance $\sigma_f = \frac{1}{\sqrt{F_{PP}}}$. The formulas above also make it clear how to remove modes in the estimator and the corresponding norm and variance calculation, which could be applied in case there exists significant contamination for those modes.

To examine the influence of the cosmological and foreground parameters, which we commonly denote as $\Theta_i$, we define the average value of the estimator following Ref. \cite{ErrorinNGs2008} 
\be
\langle \hat{a}_n \rangle  = \frac{1}{\mathcal{N}_a}\delta C^{n,\cos}_{\ell_1}\Delta^{-1}_{\ell_1\ell_2}\delta \tilde{C}_{\ell_2}
\ee
Here $\delta \tilde{C}_{\ell_2}$ is the residual $C_{\ell}$ with the true cosmological parameters. We then compute the change in $\hat{a}$ as
\be
\delta \hat{a}_n &=& \langle \hat{a}_n \rangle- a_n \nonumber\\
&=&  \frac{1}{\mathcal{N}_a}\delta C^{n,\cos}_{\ell_1}\Delta^{-1}_{\ell_1\ell_2}\delta \tilde{C}_{\ell_2}-  \frac{a_n}{\mathcal{N}_a}\delta C^{n,\cos}_{\ell_1}\Delta^{-1}_{\ell_1\ell_2}\delta C^{n,\cos}_{\ell_2} \nonumber \\
&\simeq& \frac{a_n}{\mathcal{N}_a} \delta C^{n,\cos}_{\ell_1}\Delta^{-1}_{\ell_1\ell_2}\Delta (\delta C^{n,\cos}_{\ell_2}).
\ee

%
We then conveniently define
\be
Q^i_{\hat{a}}\equiv \frac{\delta \hat{a}_n}{\delta \Theta_i} \frac{\delta \Theta_i}{a_n}\; \hspace{0.5cm} Q_{\hat{b}}\equiv \frac{\delta \hat{b}_n}{\delta \Theta_i} \frac{\delta \Theta_i}{b_n}, 
\ee
Here $\frac{\delta \hat{a}_n}{\delta \Theta_i} $ will contain the derivative of the $\delta C_{\ell}$ with respect to the cosmological parameter $\theta_i$.


As an example, let us consider $Q_{\hat{a}}$ for the variable $A_s$. After some algebra, we simply find 
\be
Q^{A_s}_{\hat{a}} = \frac{\delta A_s}{A_s},
\ee
as expected; $A_s$ is just a rescaling of the overall power and a change in this parameter should lead to a rescaling of the $a_n$ by exactly the same amount. We show the results for the other 4 $\Lambda$CDM parameters in Fig.~\ref{fig:errormodal}. $\Omega_b$ and  $H_0$ (not shown) show an increasing error for large $n$. This is because they affect the frequency of the projected signal $\delta P$, e.g. shifting the frequency can change a detection to a null signal. This does not affect the ability to capture the oscillations; instead it leads to a discrepancy between the recovered modes and the true modes, which will result in a difference between the data best-fit frequency and the mapping to that same frequency in primordial space. For $\tau$ we find a constant shift as expected. $n_s$ has very little effect on the reconstruction of the modes, with weak $n$ dependence. 

Above we have estimated the bias on the modal coefficient that results from fitting some $\delta C^{n}_{\ell}$ which were calculated for a wrong cosmology. There is however a second source of error. To fit the $\delta C^{n}_{\ell}$ to the residual power spectrum, we first have to subtract the fiducial contribution $C^0_{\ell}$ of the model without oscillations, i.e. we calculate $\hat{C}_{\ell}-C^0_{\ell}$. Since we fix the cosmological parameters in $C^0_{\ell}$, this leads to an additional error in the estimator. One can calculate the full derivative of the modal coefficient as
\be
\frac{\partial \hat{a}_n}{\partial \Theta_i} \bigg|_{\Theta_0} &=& \frac{1}{\mathcal{N}_a} \sum_{\ell} \left[ \frac{(2\ell+1)}{2} \right. \nonumber \\
&& \left. \frac{[\delta C^{n,\cos}_{\ell}]' (\hat{C}_{\ell}-C_{\ell})-\delta C^{n,\cos}_{\ell} [C_{\ell}]' }{C_{\ell}^2} \right]. 
\ee

\begin{figure}
\resizebox{1\hsize}{!}{
\includegraphics{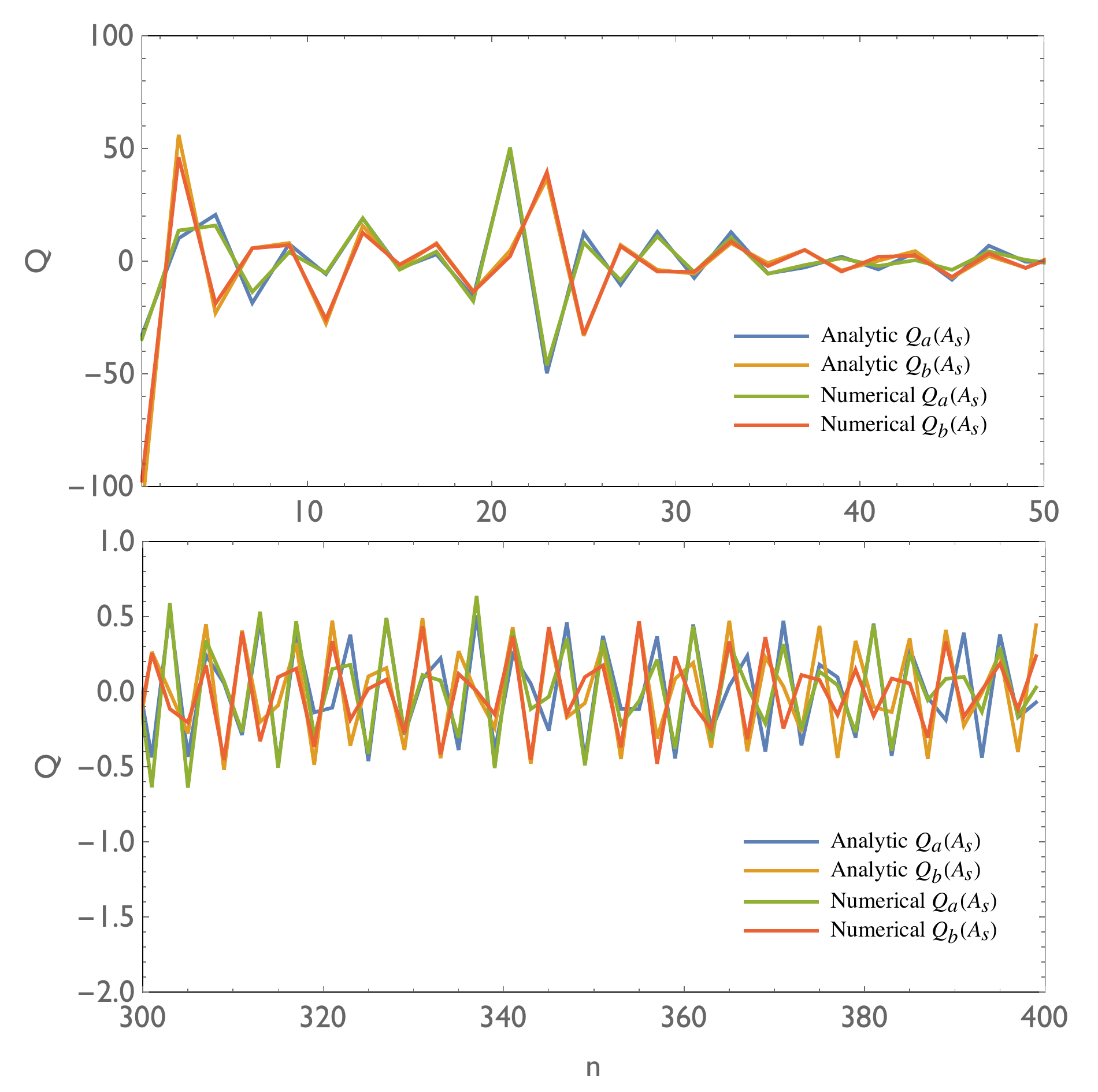}
}
\caption{Error propagation of $A_s$ into bias of $a_n$ and $b_n$. The large errors at small $n$ are expected. The lowest $n$ are a copy of the spectrum itself, while $n=20$ correspond to BAO frequencies. For large $n$ the bias is small, with only a change of half a sigma. }
\label{fig:errormodal2}
\end{figure}

 Now we take the derivatives with respect to $A_s$ (suppressing the superscripts $\cos, n$):
 \be
  \frac{\partial \hat{a}_n}{\partial A_s} = \frac{1}{A_s} \frac{1}{\mathcal{N}_a} \sum_{\ell} \left[ \frac{2\ell +1}{2}\left(-2\frac{\delta C_{\ell } C_{\ell}}{C_{\ell}^2} + \frac{\delta C_{\ell} \hat{C}_{\ell}}{C_{\ell}^2} \right)\right]
 \ee
If there is no signal $\hat{C}_{\ell} \simeq C_{\ell}$. By construction $\delta C_{\ell} /C_{\ell} < 1$. Let us denote this ratio by $r_{\ell}$. We then have 
 \be
 Q_{\hat{a}}^{A_s} \simeq - \sqrt{2}\frac{\sigma_{A_s}}{A_s} \frac{\sum (2\ell +1) r_{\ell}}{\left( \sum (2\ell +1) r_{\ell}^2 \right)^{1/2}}
 \ee
We show $Q_{\hat{a}}^{A_s}$ in Fig.~\ref{fig:errormodal2} for two ranges in $n$. We compare the analytical result to the full numerical result with $\hat{C}_{\ell}$ from Planck 2013 data \cite{Planck2013Likelihood}. It shows that in order to minimize sensitivity to errors in $A_s$, the very low $n$ should not be used. Around $n = 20-25$ there are additional large corrections, coming from the BAOs. Note that all the way to $n=50$ the wavelength of the modes is relatively large. Removing those modes should not affect the search for high frequency oscillations significantly. 
 
We will leave a full numerical treatment of the error from other cosmological parameters as well as the application of this estimator to Planck 2015 data to future work.

\subsection{Bispectrum estimator}
\label{sec:estimators_b}

\begin{figure*}
\resizebox{0.9\hsize}{!}{
\includegraphics{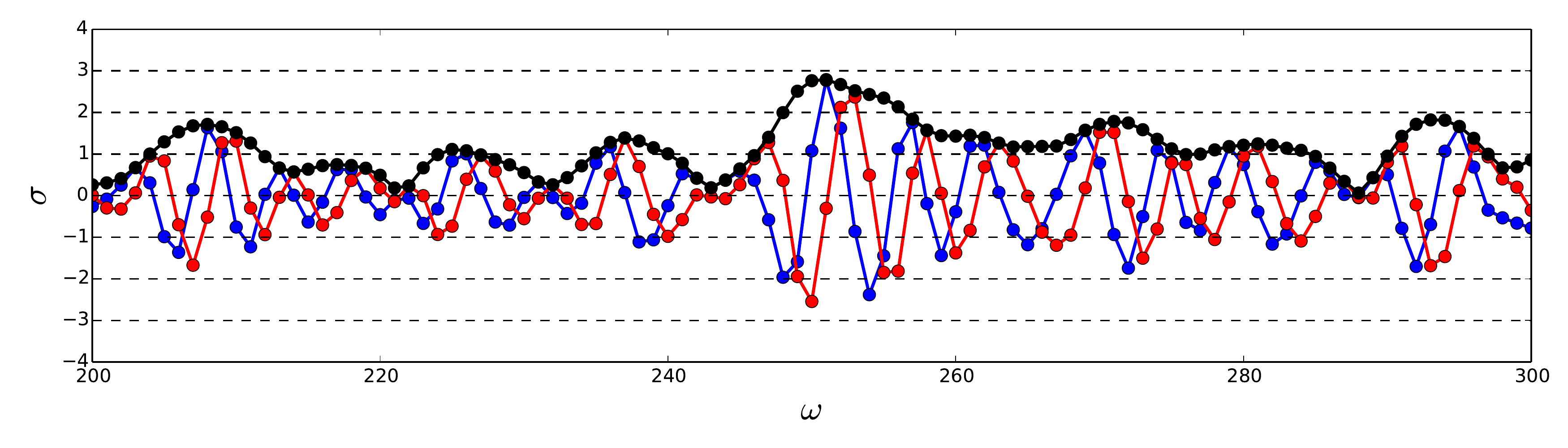}
}
\caption{Detail view of a spectrum of the bispectrum estimator. It illustrates that sine (red) and cosine (blue) component are highly correlated and therefore the full amplitude (black) has a much more stable amplitude than its components. On the y-axis, the amplitude was divided by the expected standard deviation. }
\label{fig:bispec_pdfgen2}
\end{figure*}

For the bispectrum estimator, we assume a fixed $\Lambda$CDM cosmology, and only vary the oscillation parameters. We use a KSW type optimal estimator with the effective expansion of a function $f(k_t)$ into separable linear oscillation proposed in \cite{OptimalEstimator2014},
\be
f(k_t) = \sum_{n=0}^N \left(  a_n \cos{\frac{2 \pi n k_t}{\Delta k_t}} + b_n \sin{\frac{2 \pi n k_t}{\Delta k_t}} \right).
\ee
We split the logarithmic oscillations into sine and cosine components as
\be
\label{eq_oscispectrum3} 
B^{\mathrm{feat}}_\Phi(k_1,k_2,k_3) &=& \frac{6 \Delta_\Phi^2}{(k_1k_2k_3)^2} \left[ f_1 \sin\left(\omega(k_t) \ln (k_t)\right) \right.\nonumber \\
&& \left. + f_2 \cos\left(\omega(k_t) \ln (k_t)\right)\right],
\ee
where $f_{\rm NL} = \sqrt{f_1^2 + f_2^2}$ and $\Phi = \arctan{(\frac{f_2}{f_1})}$. Here we allowed for a drifting frequency of form $\omega(k_t)$. The estimator for the two amplitude $f_1,f_2$ is then given by
\begin{equation}
\label{eq_kswphase1} 
f_i= \sum_j (F^{-1})_{ij} S_j,
\end{equation}
where $S$ is the usual sum over KSW filtered maps (see Ref.~\cite{OptimalEstimator2014} for details). The resulting bispectrum estimates from the data consist of a set of amplitudes $\hat{A}_{\omega_i,\sin}$ and $\hat{A}_{\omega_i,\cos}$ for a sufficiently tight sampling of $\omega_i$.  The overall amplitude for a given $\omega_i$ is
\begin{eqnarray}
\hat{A}^B_{\omega_i} = \sqrt{(\hat{A}^B_{\omega_i,\sin})^2+(\hat{A}^B_{\omega_i,\cos})^2}.
\end{eqnarray}
We show estimated $\hat{A}_{\omega_i,\sin}$, $\hat{A}_{\omega_i,\cos}$ and $\hat{A}^B_{\omega_i}$ in Fig.~\ref{fig:bispec_pdfgen2} for a range of frequencies, obtained from a simulated Gaussian map. On the y-axis, the estimated amplitudes have been divided by their expected standard deviation, The plot shows an interesting characteristic of logarithmic oscillations;.neighboring sine and cosine amplitudes are strongly correlated such that the overall amplitude varies much slower than the amplitude at a fixed phase. 

\begin{figure*}
\resizebox{0.98\hsize}{!}{
\includegraphics{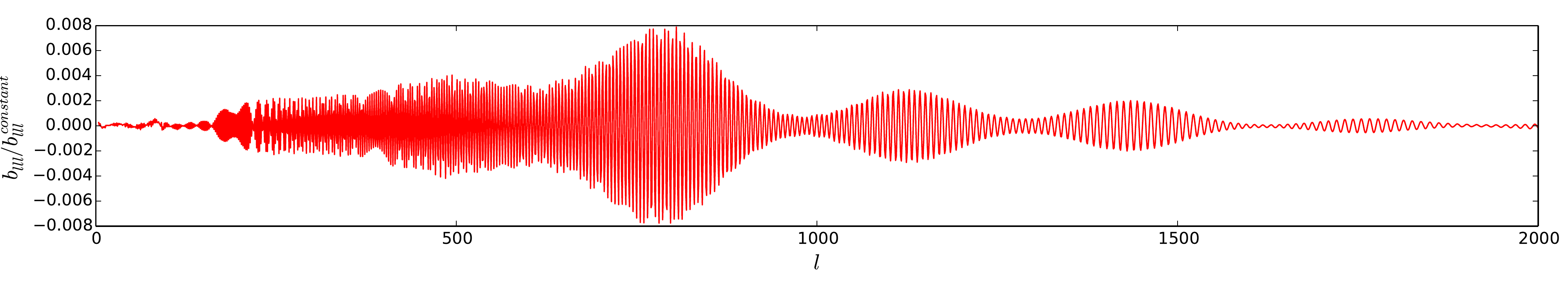}
}
\caption{Example of an ultra high frequency resonance bispectrum.}
\label{fig:logosci_examples}
\end{figure*}

With respect to  previous work, we have extended the frequency range to from $1\leq \omega \leq 1000$ (see Fig.~\ref{fig:logosci_examples} for an illustration of a very high frequency equilateral bispectrm). Our current pipeline uses 1000 sine and cosine modes. We use the high resolution version of CAMB described in Sec.~\ref{sec:estimators_p} to accurately reconstruct the high frequency modes. It is necessary to sufficiently sample the $r$ integral in the estimator; we use 1000 sampling points. This in contrast to other shapes, where often (after quadrature optimization) one can obtain good results with $\mathcal{O}(10)$ sampling points. The KSW filtered maps of the bispectrum estimator are calculated using fiducial $C_{\ell}$ that do not oscillate. We experimentally verified that using oscillating $C_{\ell}$ with an oscillation amplitude compatible with power spectrum estimates does not significantly affect the bispectrum estimator, justifying our procedure. 


\subsection{Gaussian approximation to the estimator PDF}
\label{sec:gaussianapprox}

To quantify the look elsewhere effect below, it will be extremely useful to have an analytical approximation to the PDF of the estimator. The following discussion applies to the case where cosmology and foregrounds are fixed, and only the oscillation parameters $A,\omega,\phi$ are variable. We are neglecting a possible frequency running parameter here, but it would be conceptially straight forward to include it.

The bispectrum estimated from the data consists of a set of amplitudes $\hat{A}_{\omega_i,\sin}$ and $\hat{A}_{\omega_i,\cos}$ for a sufficiently tight sampling of $\omega_i$. Our goal is to find the theoretical distribution of $\{\hat{A}^B_{\omega_i,\sin},\hat{A}^B_{\omega_i,\cos}\}$ in the case of an underlying Gaussian map, which fully describes the bispectrum estimator statistics. Note that these amplitudes are correlated. At a given frequency $\omega_i$, $\hat{A}^B_{\omega_i,\sin}$ and $\hat{A}^B_{\omega_i,\cos}$ are almost uncorrelated, but neighboring frequencies are correlated. We make no assumptions about the strength of the correlation here. Each of the two components by itself is Gaussian distributed at a given frequency $\omega_i$ with analytically calculable variance $V[A ^{\sin,\cos}_{\omega_i}]$. This is a consequence of the central limit theorem (CLT) and the large number of multipole contributions that make up a given bispectrum amplitude estimate. The common PDF of the frequency spectrum $\{\hat{A}_{\omega_i}^{\sin,\cos}\}$ of the sine and cosine components is therefore approximately a multivariate Gaussian, where the covariance matrix elements between frequencies and phases $\omega ^{\sin,\cos}_i$ and $\omega ^{\sin,\cos}_j$ can be calculated from the Fisher matrix element of the theoretical bispectrum shapes of these frequencies. We obtain a joint multivariate Gaussian PDF for the sine and cosine components
\begin{eqnarray}
P(\{\hat{A}_{\omega_i}^{\sin,\cos}\}) = \mathcal{N}(\mu=0, \Sigma),
\end{eqnarray}
with $\Sigma=\frac{F_{ij}}{F_{ii}F_{jj}}$, where $F_{ij}$ is the corresponding Fisher matrix. Of particular interest is the distribution of the overall amplitudes $P(\{\hat{A}_{\omega_i}\})$ which allows us to obtain the PDF for the maximum $P(\bar{A}_i^{\text{max}}\ge x)$. While it may be possible to make analytic progress, it is easy to sample $P(\{\hat{A}_{\omega_i}\})$ with fast sampling from the underlying multivariate Gaussian. The sampling strategy also allows to estimate the combined look elsewhere effect in the next section. Generating mock bispectrum amplitude spectra in this way is significantly  faster than analyzing Monte Carlo map simulations.

In the case of the bispectrum, the Fisher matrix between frequencies $\{\omega_i,\omega_j\}$ and phases $\{\sin,\cos\}_{i,j}$ is given by 
\be
\label{eq_fisher} 
F_{ij} = \sum_{\ell_1\ell_2\ell_3}  \frac{ B^i_{\ell_1\ell_2\ell_3} B^j_{\ell_1\ell_2\ell_3} }{C_{\ell_1}C_{\ell_2}C_{\ell_3}},
\ee
which can be evaluated from the mode function Fisher matrix $F^{mode}_{nm}$ as
\begin{align}
\label{eq_fisher4} 
F_{BB'} = \mathbf{a_n F^{mode}_{nm} a_m}.
\end{align}
In the presence of noise, the $C_l$ are modified ad $C_l \rightarrow C_{\ell} + N_{\ell}$ where $N_{\ell}$ is the isotropic noise power spectrum. Partial sky coverage approximately leads to $F_{ij} \rightarrow f_{\rm sky} F_{ij}$ where $f_{\rm sky}$ is the covered sky fraction. The resulting bispectrum amplitude samples include homogeneous noise and partial sky coverage. Instrumental effects that are not captured in this way cannot be represented by the method proposed here. We argue that this is not a concern for bispectrum measurements after the underlying maps have been appropriately cleaned. 

As explained above, for the power spectrum, one can sample the PDF by generating large amounts of simulated maps. However, in analogy to the bispectrum, for a fixed cosmology, an analytic approximation to the PDF can be obtained from the power spectrum Fisher matrix. We assume that the estimate is of form $C_{\ell} = C_{\ell}^{\rm fid} + \hat{A} C_{\ell}^{\rm osc}$ i.e. a fixed fiducial spectrum $C_{\ell}^{\rm fid}$. The Fisher matrix is then $F =C^{\rm{osc}}_{\ell_1}\Delta^{-1}_{\ell_1\ell_2}C^{\rm{osc}}_{\ell_2}$ where $\Delta_{\ell_1\ell_2}$ is the spectrum covariance matrix of the experiment. If the off diagonal elements of the covariance matrix are negligible, which we assume for the simulations below, then the Fisher matrix reduces to
\be
\label{eq_fisher2} 
F_{ij} = \sum_{l} \frac{ C^{i,{\rm osc}}_{\ell} C^{j,{\rm osc}}_{\ell} }{(\Delta C_{\ell})^2}.
\ee
In the cosmic variance limited case
\be
\Delta C_{\ell} = \sqrt{\frac{2}{2\ell+1}} C_{\ell}^{\rm fid}.
\ee

\begin{figure*}
\resizebox{0.9\hsize}{!}{
\includegraphics{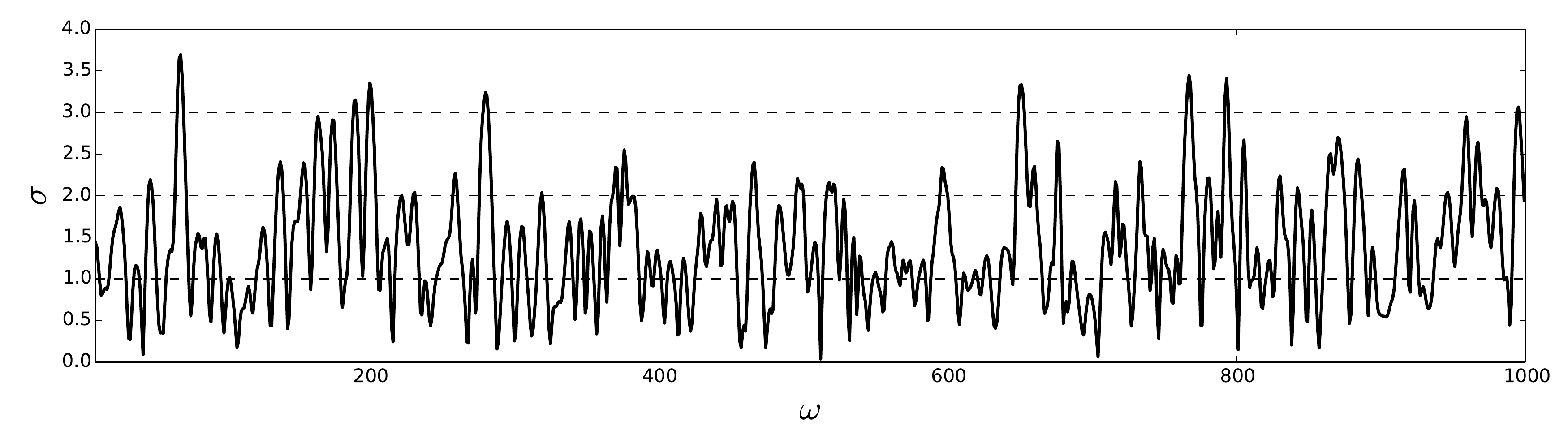}
}
\resizebox{0.9\hsize}{!}{
\includegraphics{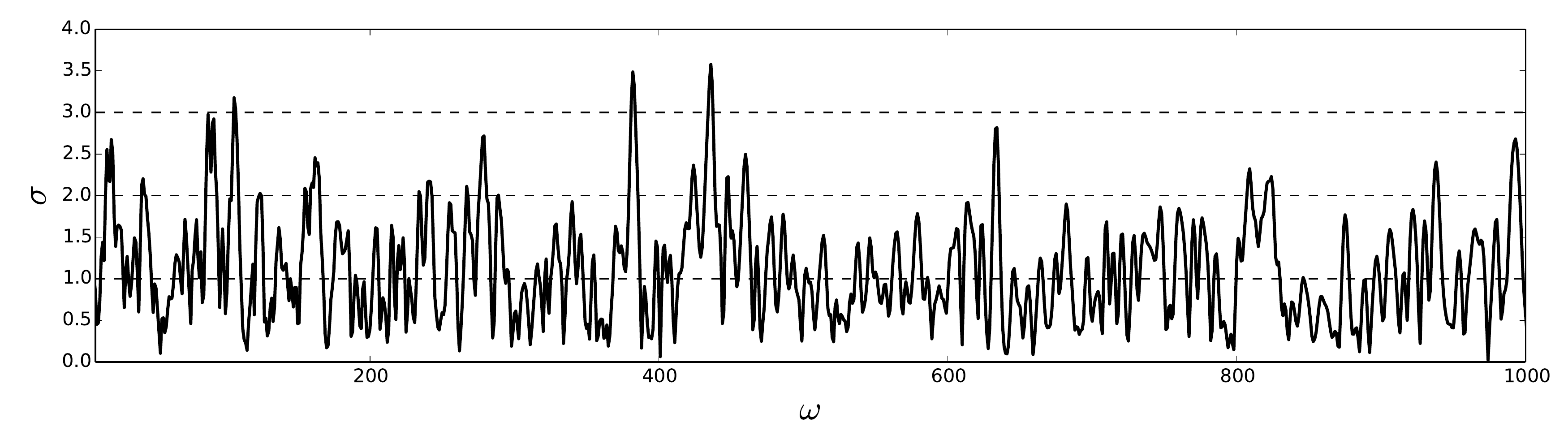}
}
\caption{Top: A spectrum of the bispectrum estimator drawn from the bispectrum correlation matrix. Bottom: A spectrum from a simulated Gaussian map.}
\label{fig:bispec_pdfgen1}
\end{figure*}
\begin{figure*}
\resizebox{0.9\hsize}{!}{
\includegraphics{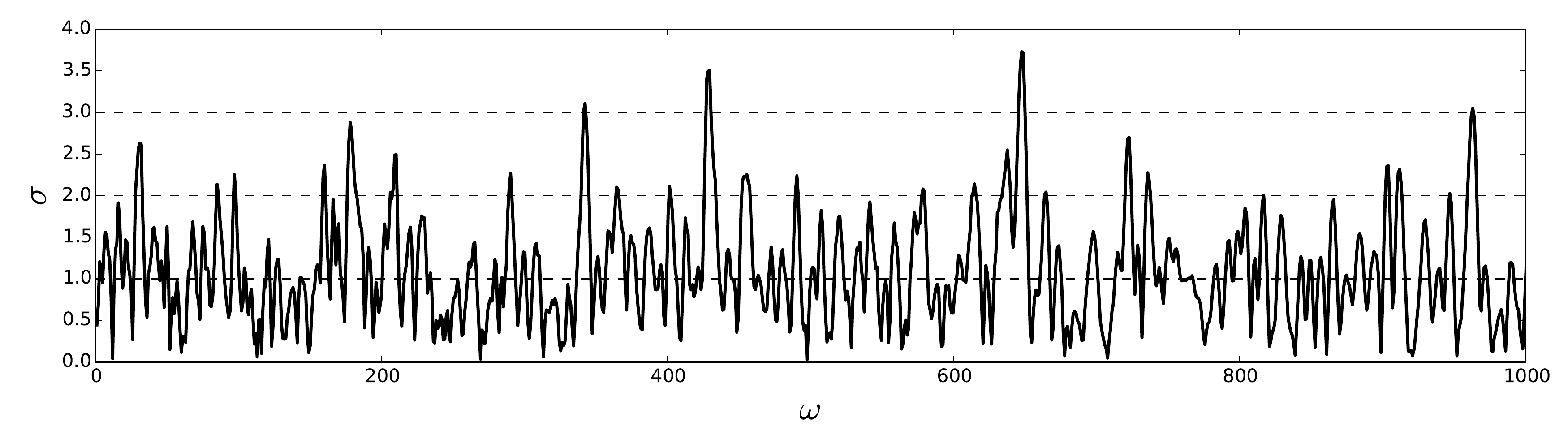}
}
\caption{A spectrum of estimated amplitudes drawn from the correlation matrix of the power spectrum. Note that this plot is for an idealized experiment with a diagonal covariance matrix, as Planck has not released the spectrum covariance matrices independently. Our method also applies to a realistic covariance matrix as in the case of the Planck experiment. The off-diagonal coupling in this case leads to decreased frequency resolution.}
\label{fig:powspec_pdfgen1}
\end{figure*}

Examples of spectra generated in this way are shown in Fig.~\ref{fig:bispec_pdfgen1} for the bispectrum and in Fig.~\ref{fig:powspec_pdfgen1} for the power spectrum. We also generated a spectrum that results from a simulated map. The full simulation matches the structure of the analytic PDF spectrum, with peaks that are statistically of the same width and height, as expected. We also see that power spectrum and bispectrum have a similar frequency resolution. Note however that the power spectrum plot is generated with a diagonal covariance matrix. This is not a good approximation for a realistic experiment. Using a realistic covariance matrix for the Planck experiment (not shown), one finds that the frequency resolution is somewhat reduced by the off diagonal covariance matrix terms.


\section{Combined estimator statistics and look-elsewhere effect}
\label{sec:combined}

In this section we explain how to define estimators combining the power spectrum and the bispectrum, as well as how to calculate significances that take into account the look-elsewhere effect.


\subsection{Statistical independence of bispectrum and power spectrum}
\label{sec:statindep}

The estimators for the power spectrum and bispectrum are based on the same data maps, and are therefore a priori correlated. This correlation in the data exists already for a Gaussian map. In addition, if the underlying model truly contains an oscillating signal, then the correlation between amplitudes and phases contains additional information to test a given model. We will now argue that the power spectrum and bispectrum in the Gaussian assumption are to good approximation statistically independent. We then use this property to find optimal combined estimators and quantify the look elsewhere effect in the following sections. 

A high degree of statistical independence follows from the following simple argument. By themselves, power spectrum and bispectrum amplitude estimates at a given phase are closely approximated by a gaussian PDF, since they sum over a large number of random variables (CLT). The correlator of the power spectrum and bispectrum amplitude is a sum over 5-point functions of $a_{lm}$ which are zero in a Gaussian map, as they cannot be expressed in terms of two point functions. Therefore the joint PDF of power spectrum and bispectrum amplitude estimates factorizes into two statistically independent multivariate Gaussians, as described in the previous section.

The CLT conditions are not fully satisfied as we have finite $\ell_{\rm max}$. The true joint PDF is therefore not exactly Gaussian and the fact that they are statistically uncorrelated does not imply statistical independence. The authors of of Ref.~\cite{ShellardBispectrumPsJoined2014} give an analytic argument that the true correlation of power spectrum and bispectrum amplitude estimates $\rm{Corr}(|\hat{A}_P|,|\hat{A}_B|)$ should be of order $\frac{1}{\ell_{\rm max}}$, independent of the shape under consideration. This is sufficiently small to be neglected in this study and from now we assume statistical independence of the power spectrum and bispectrum.

\subsection{Combining power spectrum and bispectrum}

We now discuss how to combine the power spectrum and bispectrum analysis, in the absence of a full  bispectrum likelihood over cosmological parameters and foregrounds.

\subsubsection*{Case 1: fixed power spectrum and bispectrum cosmology}

The bispectrum analysis is schematically done as follows: Each bispectrum shape $S_B(\omega,\Phi,...)$ is defined by a number of parameters, the frequency $\omega$ and phase $\Phi$ and possibly some other parameters. The bispectrum estimator gives a best fit amplitude $\hat{A}^S_B$ to this shape. To combine estimators with the power spectrum, we want power spectrum amplitude estimates $\hat{A}^S_P$ for the corresponding power spectrum shape $S_P(\omega,\Phi,...)$. As discussed before, a fair way to combine the two analyses is to also fix the cosmology in the power spectrum analysis to its best fit value without oscillations. In that case the power spectrum likelihood has an analytic ML estimator, as described in~\cite{ShellardBispectrumPsJoined2014}, which is given by a simple quadratic fit of the oscillating power spectrum shape to the measured $\hat{C}_l$:
\begin{eqnarray}
\hat{A}^S_P&=&\frac{1}{N} C^{\rm{osc}}_{\ell_1}\Delta^{-1}_{\ell_1\ell_2}\left(\hat{C}_{\ell_2}-C^{\rm{fix}}_{\ell_2}\right),\\
N_P&=&C^{\rm{osc}}_{\ell_1}\Delta^{-1}_{\ell_1\ell_2}C^{\rm{osc}}_{\ell_2}.
\end{eqnarray}
The estimator is approximately Gaussian and its distribution can be calculated from the Fisher matrix as we described in the previous section. In some sense this procedure gives a model the best chance to be significant, since there is no look elsewhere effect associated with cosmological and foreground parameters that could dilute significance. Our quadratic power spectrum modal estimator proposed in Sec.~\ref{sec:estimators_modal} is of this form.

\subsubsection*{Case 2: likelihood for the power spectrum, fixing cosmology for the bispectrum}

As we explained in Sec.~\ref{sec:estimators_p}, the power spectrum search for oscillations in usually done by exploring a likelihood $\mathcal{L}(c,f,o)$ over cosmological $(c)$, foreground $(f)$ and oscillation $(o=\{A_P,\omega,\phi\})$ parameters. One then obtains an effective $\chi^2$ by computing
\be
\chi^2_{\rm{eff}} = 2 \left(\ln \mathcal{L}_{\rm{osc}} - \ln \mathcal{L}_{\rm{base}} \right),
\ee
which can be compared to the $\chi^2$ distribution with the associated number of degrees of freedom. To correct for the look elsewhere effect one can follow a frequentist approach and compare the measured $\chi^2$ to the distribution of $\chi^2_{\rm{eff,max}}$ from Monte Carlo map simulation. One may also compute the Bayes factor from the likelihood. 

There are two related ways to calculate combined estimators with the bispectrum, either by translating the power spectrum estimate to the language of amplitudes or the bispectrum estimate to the language of likelihoods. For the bispectrum we are estimating the amplitude $A_B(\omega,\phi)$ for each set of oscillation parameters $\omega,\phi$. We may therefore calculate power spectrum ML amplitude estimates $A^{\rm{ML}}_P(\omega,\phi)$ by maximizing the likelihood $\mathcal{L}_P(f,o,A_p,\omega,\phi)$ with respect to foreground $(f)$ and oscillation $(o)$ parameters as well as amplitude $A_P$ for each set $(\omega,\phi)$. The amplitudes $A^{\rm{ML}}_P(\omega,\phi)$ and $A_B(\omega,\phi)$ can then be used to obtain combined estimators. Note that it is necessary to know the variance of $A^{\rm{ML}}_P(\omega,\phi)$, which unlike in the previous section cannot be calculated from the Fisher matrix but has to be obtained from a Monte Carlo excursion. 

Based on the independence of power spectrum and bispectrum data, we can also write down a simple combined likelihood. The bispectrum likelihood only contains the oscillation parameters and is evaluated for a fixed cosmology. The combined likelihood is then
\be
\ln \mathcal{L}(c,f,\omega,A,\phi) = \ln \mathcal{L}_P(c,f,A_P,\omega,\phi) + \ln \mathcal{L}_B(A_B,\omega,\phi), \nonumber \\ 
\ee
where $\mathcal{L}_P$ is the usual power spectrum likelihood. The bispectrum likelihood $\mathcal{L}_B $ is a Gaussian distribution for the measured sine and cosine amplitudes $\{\hat{A}_{\omega_i}^{\sin},\hat{A}_{\omega_i}^{\cos}\}$, whose mean and covariance matrix is determined by the model parameters $(A_B,\phi)$, as described in more detail in Sec. 4 of~\cite{LinearOscillationsMoritz2014}.



In the following, we will use the language of amplitudes to define combined estimators, since it relates more directly to the form of the shapes in Sec.~\ref{sec:models}. 



\subsection{Generalities about peak estimators and look-elsewhere effect}

\subsubsection{Peak estimators}

In this section we define look-elsewhere corrected peak estimators for the models reviewed in section \ref{sec:models}. By peak estimators we mean that the most significant amplitude(s) is picked from the spectrum of the estimator by maximizing over its variables (e.g. frequency and phase). The starting point of all peak estimators in this section are the amplitude estimates $\hat{A}_{X}(\omega,\phi)$, where X=P is the power spectrum and X=B is the bispectrum. These are to good approximation Gaussian distributed with mean zero. The variance of these estimators can be calculated from the Fisher matrix. Different phases are not independent and in the case of the bispectrum we start from the sine and cosine amplitudes $\hat{A}_{X}(\omega,\phi=0),\hat{A}_{X}(\omega,\phi=90^\circ)$, which determine the value for any phase $\phi$.

Because of the different questions that one can pose to the data, which lead to different probabilities, it is particularly important to adopt to a consistent notation scheme. We write any amplitude estimator $\hat{A}$ with a hat, with a subscript listing what the estimator depends on (the free variables) and a superscript listing the variables that have been maximized over. For example 
\be
\hat{A}_{P,\omega}^{\rm{max}(\phi)} = \max_\phi \hat{A}_{P}(\omega,\phi),
\ee
is the amplitude estimate from power spectrum at a given frequency $\omega$, maximized over the phase $\phi$.  

For Gaussian random variables with zero mean (for example the oscillation amplitude at a fixed phase) we define the significance (the number of $\sigma$'s) of the estimate with a bar as
\begin{equation}
\label{eq:significance_gauss}
\bar{A}=\frac{\hat{A}}{\text{Var}\left[\hat{A}\right]^{\frac{1}{2}}}.
\end{equation}
For general random variables, we define the significance $\bar{A}$ as 
\begin{equation}
\label{eq:significance2}
\bar{A}=\sqrt{2} \mathrm{Erf}^{-1}(1-p(\hat{A})),
\end{equation} 
where $p(\hat{A})$ is the corresponding p-value of the estimate, i.e.~we transform p-values to significances of a Gaussian distribution, as was done in Refs.~\cite{PlanckNGs2015} and \cite{ShellardBispectrumPsJoined2014,Fergusson:2014tza}.

\subsubsection{Sampling the PDFs}

The significance for any CMB peak estimator can be obtained from its PDF in the case of an underlying Gaussian CMB map. For example, the p-value of a measurement of $\hat{A}_{X,\omega}^{\rm{max}(\phi)}$ follows from the PDF 
\be
P\left( \hat{A}_{X,\omega}^{\rm{max}(\phi)} \right).
\ee
This p-value takes into account the look-elsewhere effect (in this example generated from the maximization over the phase). 

A key point of this analysis is to obtain PDF's of the estimators considered in this paper, which may for example combine power spectrum and bispectrum amplitudes at multiple frequencies. The usual method to obtain this distribution is from simulated Gaussian maps.  However, the high resolution bispectrum estimators cannot easily be run on a large number of maps because of its high computational costs. It is possible to find analytic approximations for the peak estimator PDF \cite{ShellardBispectrumPsJoined2014}, but these usually have to be calibrated by simulated maps. 

Here we propose a new method to obtain the distribution of estimators derived from the $\hat{A}_{X,\omega,\phi}$. We sample spectra of estimators not from Gaussian maps but from their theoretical PDF discussed in section \ref{sec:gaussianapprox}. This can be done very rapidly once the Fisher matrix is calculated. As we explained in Sec.~\ref{sec:gaussianapprox}, we expect this to be a good approximation of the data, even though inhomogeneous noise cannot be taken into account. In particular, we assume only Gaussianity of the estimated amplitudes, and not e.g. Gaussianity of the underlying $a_{\ell m}$. In the following section we evaluate the peak estimator PDFs for resonance models using this technique. The corresponding PDFs for linear feature models have previously been studied nicely in \cite{ShellardBispectrumPsJoined2014}.


\subsection{Single frequency peak estimators}

\subsubsection{Single frequency power spectrum or bispectrum survey}

\begin{figure*}
\resizebox{0.8\hsize}{!}{
\includegraphics{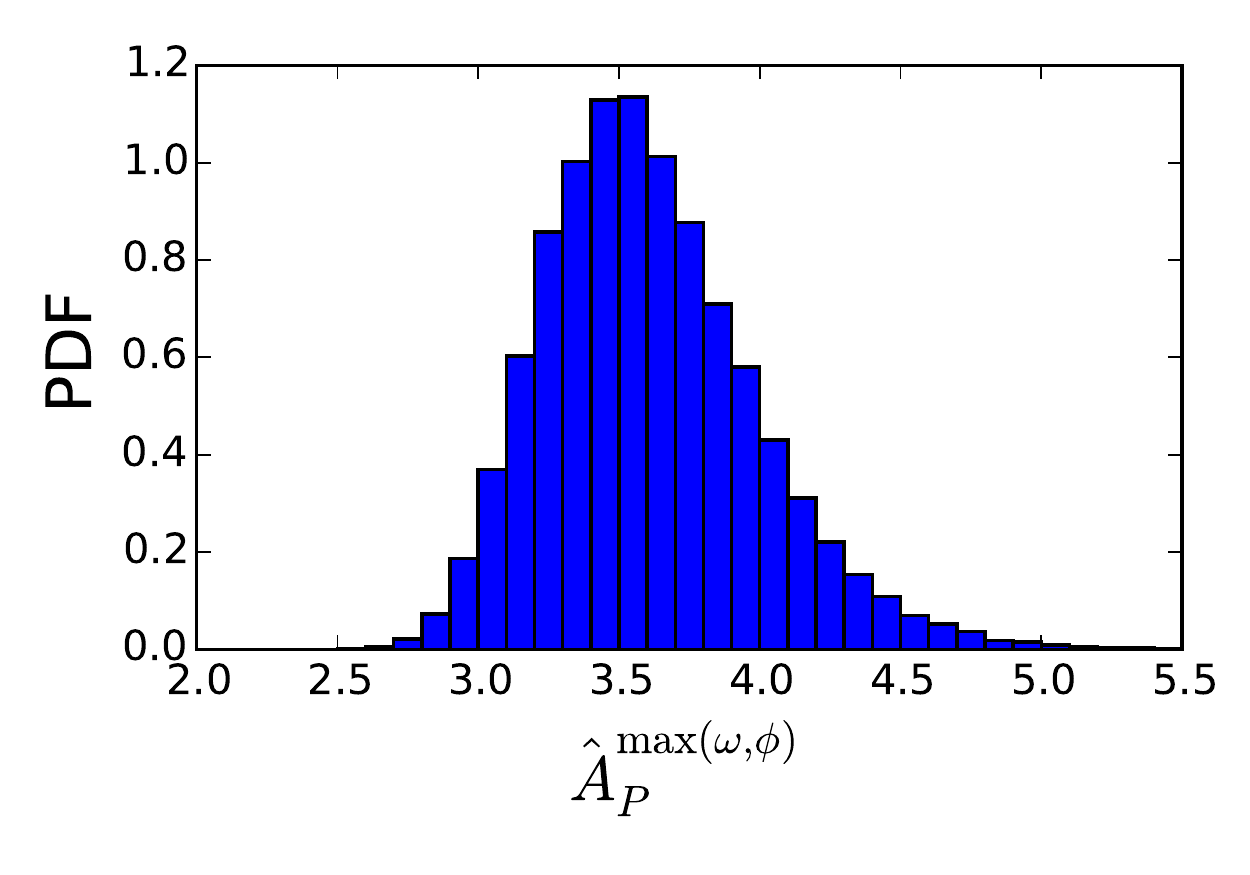}
\includegraphics{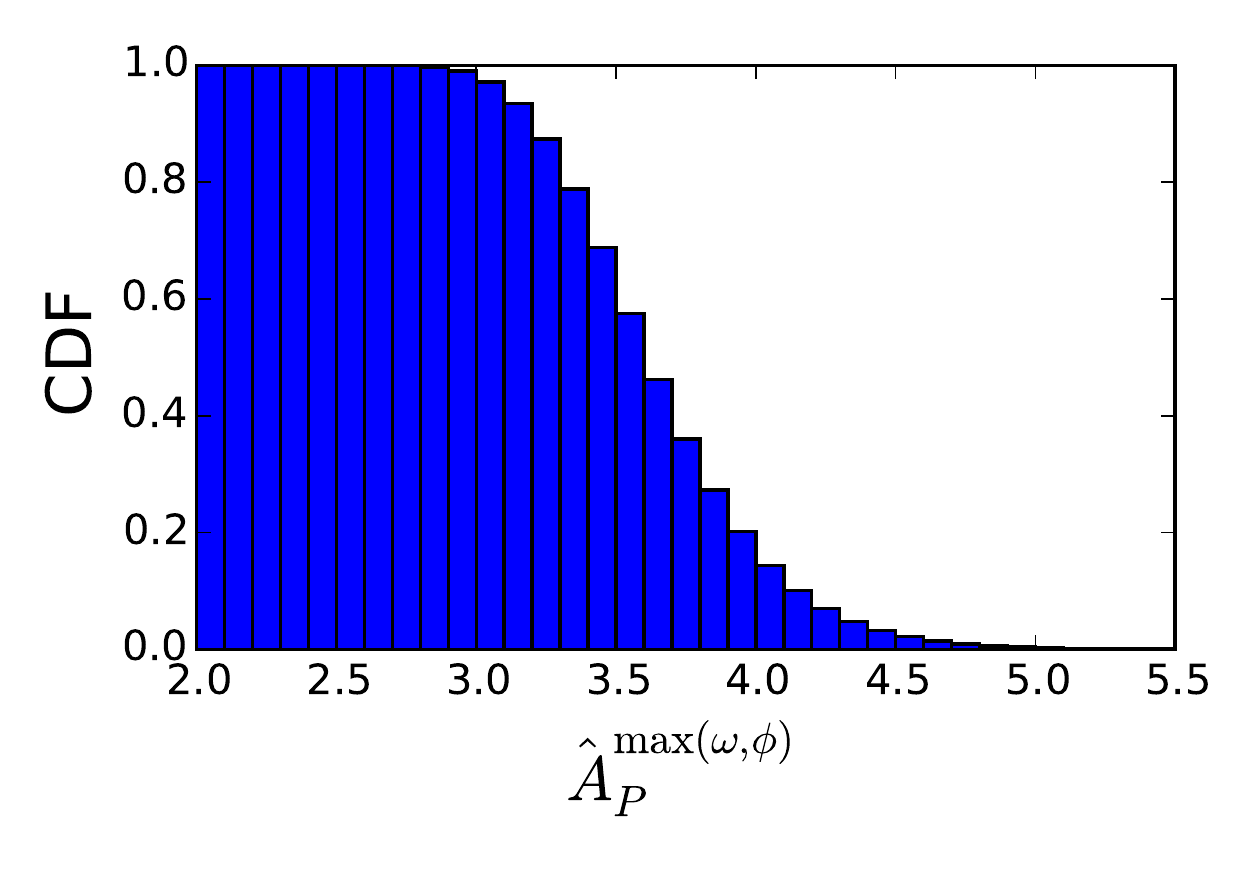}
}
\resizebox{0.8\hsize}{!}{
\includegraphics{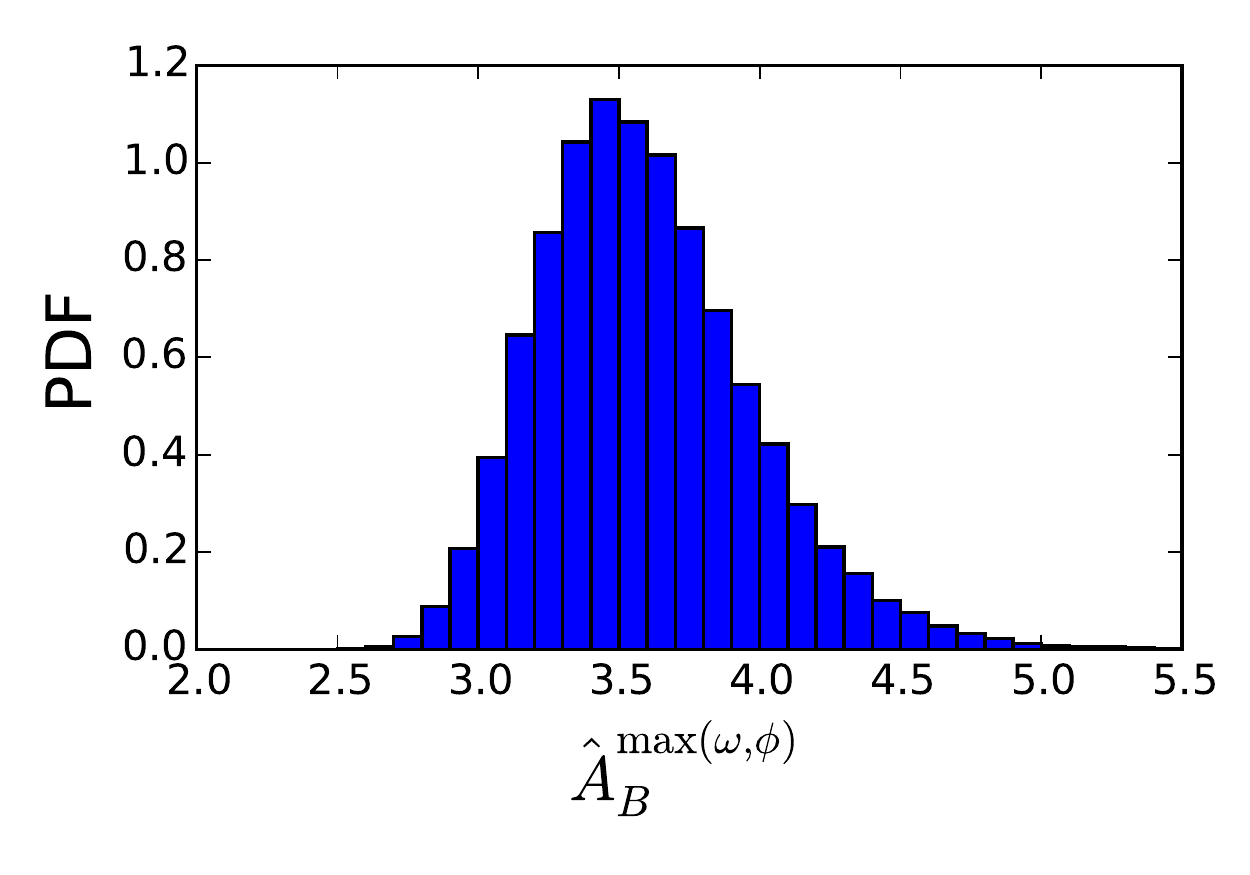}
\includegraphics{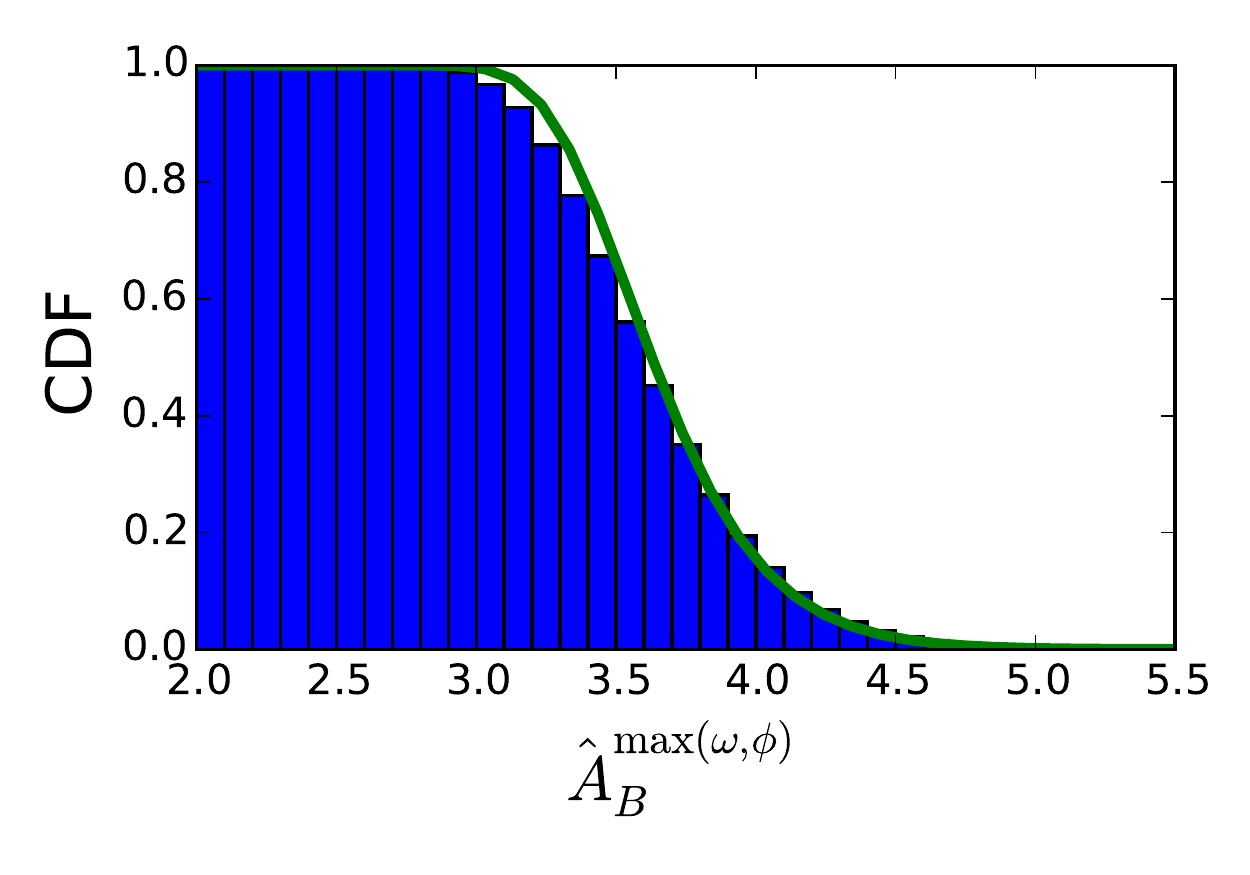}
}
\caption{Left: The PDF and CDF of the single frequency estimator Eq.~\eqref{eq:amax_single} for a power spectrum (top) and bispectrum (bottom) survey with with $0<\omega<1000$. These histograms where generated from 50.000 Gaussian estimator spectra obtained with the Fisher matrix method. The bispectrum PDF plot also shows the analytic approximation of ~\cite{ShellardBispectrumPsJoined2014} with $N_{\rm{eff}}=500$. The present plots are for an experiment with $30 < \ell < 2000$ with a diagonal cosmic variance covariance matrix. The generalisation to a realistic off-diagonal covariance matrix is straight forward.}
\label{fig:amax_single}
\end{figure*}

We start with the simple case of searching for the largest excess in the frequency spectrum of either power spectrum or bispectrum, with a free phase $\phi$. We thus have the single peak estimator
\be
\label{eq:amax_single}
\hat{A}_{X}^{\rm{max}(\omega,\phi)} &=& \max_{(\omega,\phi)} \bar{A}_{X}(\omega,\phi)\\
&=& \max_{(\omega)} \sqrt{ \bar{A}^2_{X}(\omega,\phi=0) + \bar{A}^2_{X}(\omega,\phi=90^\circ) },\nonumber \\
\ee
where the $\bar{A}_{X}(\omega,\phi)$ are the significances of the amplitudes obtained from the data and normalized by Eq.~\eqref{eq:significance_gauss}. The analytic maximization over $\phi$ in the second equation is only valid when the sine and cosine amplitudes at a given frequency are independent and have the same variance (both of which is approximately true but not required for this analysis). As pointed out in the previous section, to calculate the significance of the single peak estimate, we need the probability density function $P\left( \hat{A}_{X}^{\rm{max}(\omega,\phi)} \right)$ for Gaussian maps. An example of the PDF of Eq.~\eqref{eq:amax_single} for bispectra measured with Planck is shown in Fig.~\ref{fig:amax_single}. It was obtained by sampling 50.000 bispectrum amplitude spectra from their estimator PDF obtained in section \ref{sec:gaussianapprox}.

An analytic approximation for the cumulative distribution function is given in Ref.~\cite{ShellardBispectrumPsJoined2014}, 
\begin{eqnarray}\nonumber
\label{eq:cdf_single}
&\quad&P(\hat{A}_{X}^{\rm{max}(\omega,\phi)} \ge x)=1-\left(1-\exp{\left(-\frac{x^2}{2}\right)}\right)^{N_{\rm eff}},
\end{eqnarray}
where $N_{\rm eff}$ is the effective number of independent frequencies $\omega$ and is taken from a fit to the Monte Carlo results. As discussed before, true Monte Carlo on map basis is not computationally possible in our case, however we can again use the Fisher matrix to draw samples and determine $N_{\rm eff}$. We show a comparison of the CDF obtained with our method and of the analytic approximation \eqref{eq:cdf_single} in Fig.~\ref{fig:amax_single}, and find a best fit for $N_{\rm eff}=500$ assuming $0<\omega<1000$.





\subsubsection{Single frequency combined power spectrum and bispectrum survey}

In general, the minimal variance estimator for an amplitude $\hat{A}$, given two independent measurements $\hat{A}_P$ and $\hat{A}_B$ of a Gaussian random variable, is given by the inverse variance weighting
\begin{eqnarray}
\hat{A}_{PB}&=&\frac{1}{N}\left(\frac{\hat{A}_P}{V_P}+\frac{\hat{A}_B}{V_B}\right),\\
N&=&\frac{1}{V_P}+\frac{1}{V_B},
\end{eqnarray}
with variance $\<\hat{A}^2 _{PB}\>=\frac{1}{N}$. Here it was assumed that bispectrum and power spectrum are normalized to have the same amplitude. If one introduces an amplitude ratio $r=A_B/A_P$, then for a given $r$ the optimal estimator is \cite{ShellardBispectrumPsJoined2014}
\begin{equation}
\hat{A}_{PB}=\frac{\hat{A}_P/V_P+r\hat{A}_B/V_B}{V_P^{-1}+r^2V_B^{-1}}.
\end{equation}
So far we have not been precise about which quantities are to be combined in the power spectrum and the bispectrum. All the models under consideration predict $\omega_P=\omega_B$, so we will only combine same frequency amplitudes. For the amplitude ratio $r$ and phase difference $\Delta \phi$ there are two possible options; keeping them fixed by a model prediction, or varying them. It is important to be precise about this issue when estimating the look elsewhere effect, since more free parameters give more look-elsewhere effect.

\begin{figure*}
\resizebox{0.8\hsize}{!}{
\includegraphics{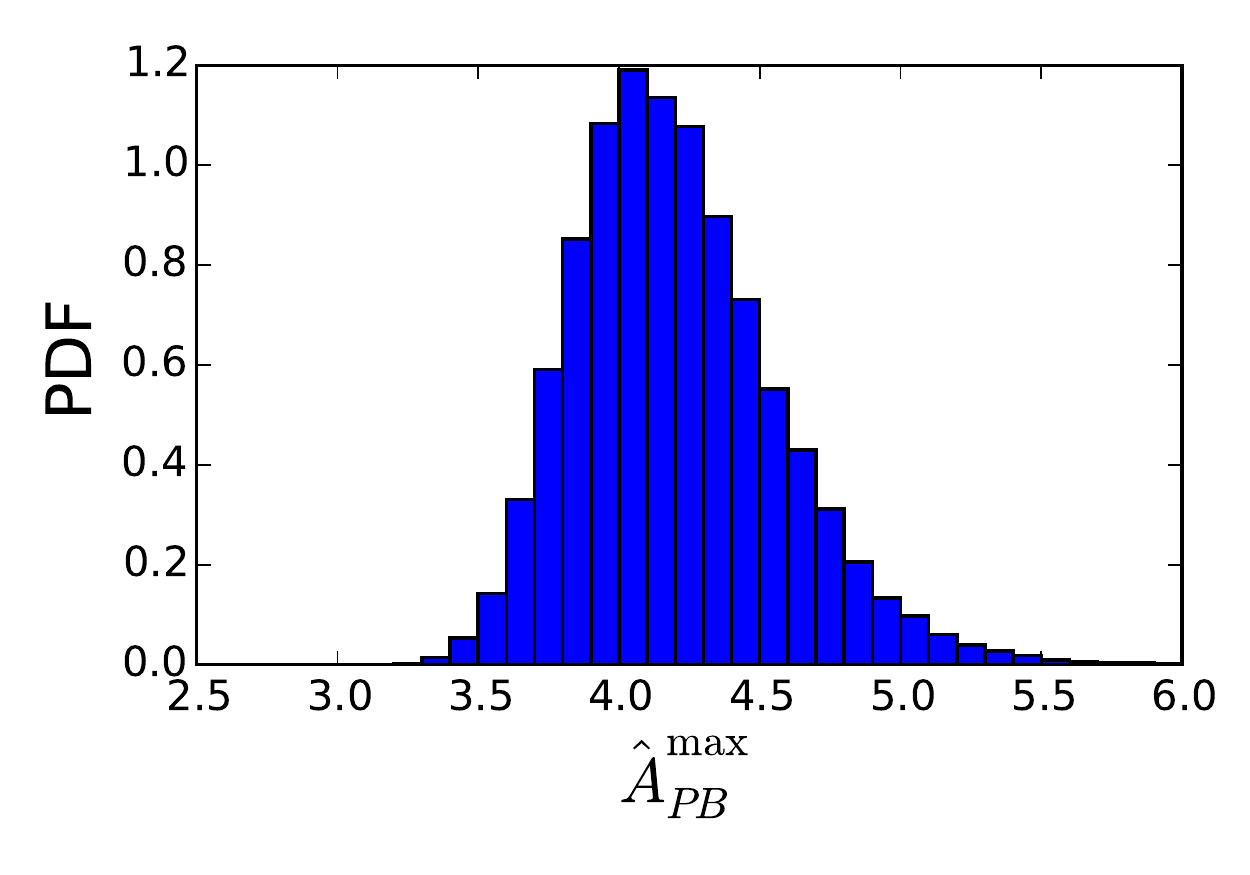}
\includegraphics{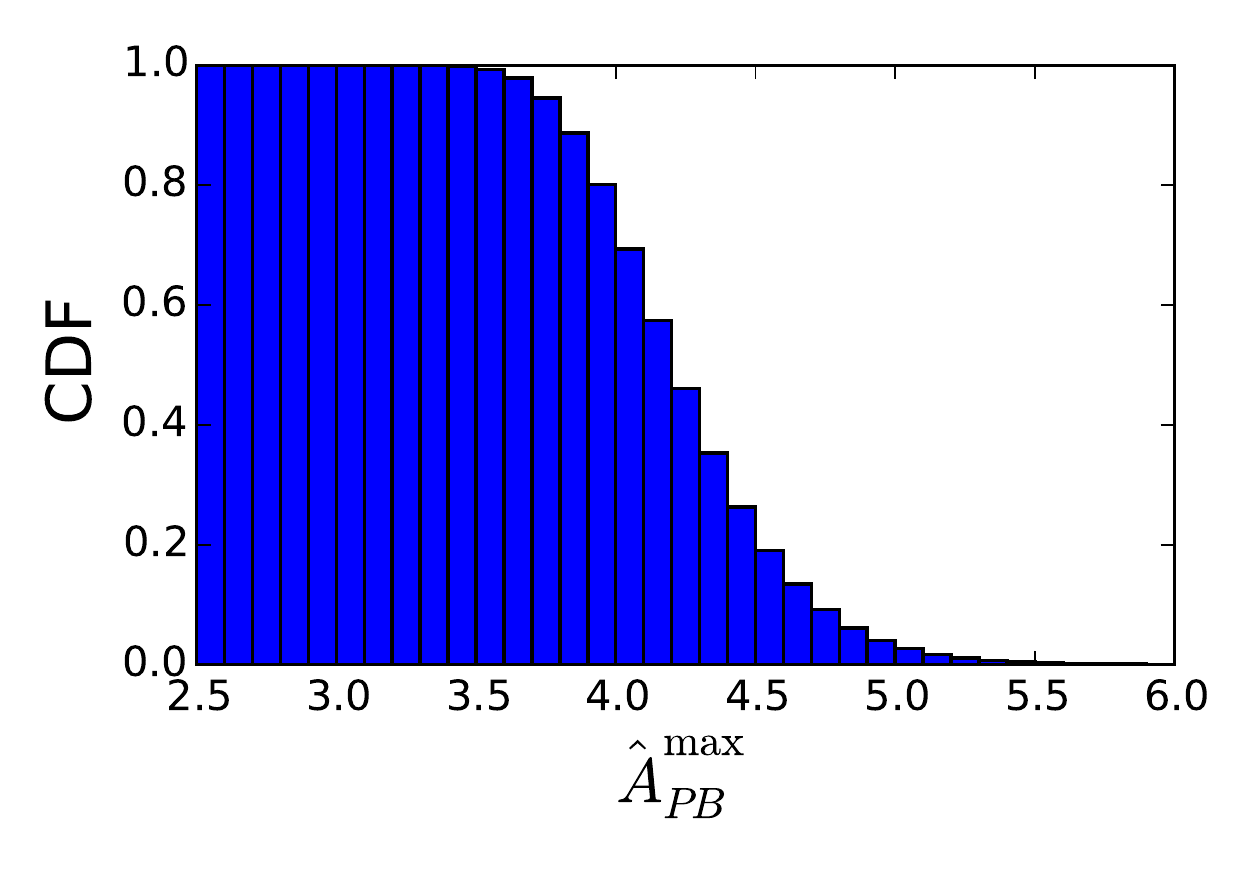}
}
\caption{Left: The PDF and CDF of the combined P plus B single frequency estimator Eq.~\eqref{eq:estim_comb1} for a survey with with $0<\omega<1000$. These histograms where generated from 50.000 Gaussian estimator spectra obtained with the Fisher matrix method for an experiment with $30 < \ell < 2000$ with a diagonal cosmic variance covariance matrix.}
\label{fig:amax_single_combi}
\end{figure*}

When combining power spectrum and bispectrum, we can consider several possibilities. The most general search leaves the phase differences and relative amplitude $r$ open. The estimator of interest is
\be
\label{eq:estim_comb1}
\hat{A}_{PB}^{\rm{max}(r, \phi_P, \phi_B, \omega)} &=& \max_{r, \phi_P, \phi_B, \omega} \bar{A}_{PB} (\omega, \phi_P, \phi_B, r)\\
&=& \max_{\omega} \sqrt{ \left( \bar{A}_{P,\omega}^{\rm{max}(\phi_P)}\right)^2 + \left( \bar{A}_{B,\omega}^{\rm{max}(\phi_B)}\right)^2 },\nonumber \\
\ee
where the second equation uses the fact that bispectrum and power spectrum measurements are independent. More constraining is a fixed model-given amplitude ratio $r$ such that
\begin{equation}
\hat{A}_{PB}^{\rm{max}(\phi_P, \phi_B, \omega)} = \max_{\phi_P, \phi_B, \omega} \bar{A}_{PB} (\omega, \phi_P, \phi_B),
\end{equation}
or a fixed model given $\Delta \phi$ (so that $\phi = \phi_B = \phi_P + \Delta \phi)$
\begin{equation}
\hat{A}_{PB}^{\rm{max}(r, \phi, \omega)}  = \max_{r, \phi, \omega} \bar{A}_{PB} (r, \omega, \phi),
\end{equation}
or a combination of both
\begin{equation}
\hat{A}_{PB}^{\rm{max}(\phi, \omega)} = \max_{\phi, \omega} \bar{A}_{PB} (\omega, \phi).
\end{equation}
The PDFs of all these estimators can be obtained by sampling the theoretical distribution of the estimator for a Gaussian map. From these distributions one can then calculate the look-elsewhere effect corrected p-value of an estimate $\hat{A}$. From the p-value one gets the equivalent significance $\bar{A}$ using Eq.~\eqref{eq:significance2}. As an example, we plot the estimator Eq.~\eqref{eq:estim_comb1} in Fig. \ref{fig:amax_single_combi}.

\subsubsection{Running frequency and further free parameters}

Similarly we can consider shapes that have additional free parameters. Foe example, we consider a frequency running due to an exponent $p_f$ as described in Sec.~\ref{sec:models}. The amplitude estimates obtained from the data are $\hat{A}_{X}(\omega,\phi,p_f)$. The single peak estimator from power spectrum or bispectrum is then
\begin{equation}
\hat{A}_{X}^{\rm{max}(\omega,\phi,p_f)} = \max_{(\omega,\phi)} \bar{A}_{X}(\omega,\phi,p_f),
\end{equation}
which again has to be sampled to obtain $P\left( \hat{A}_{X}^{\rm{max}(\omega,\phi,p_f)} \right)$ to assign significances. Sampling this probability density function requires to calculate the Fisher matrix of shapes with $p_f$ within a sufficient prior range.

\subsection{Multi frequency peak estimators}

Next, we generalize our discussion to the multi-frequency models of section \ref{sec:models}. We could for example consider the $M$ highest peaks found in the data. We assume that the amplitude ratios and phases are free parameters and drop their indices to simplify notation. We are interested in the multi-peak amplitude estimator
\begin{equation}
\label{eq:multi_single}
\hat{A}_{X,M} = \left( \sum_{i=1}^M \bar{A}_{X,i}^2 \right)^{1/2}, 
\end{equation}
for an independent power spectrum or bispectrum measurement or
\begin{equation}
\hat{A}_{PB,M}= \left( \sum_{i=1}^M \left( \bar{A}_{P,i}^2 + \bar{A}_{B,i}^2 \right) \right)^{1/2},  
\end{equation}
for a combined measurement. The $i$ index presents the $M$ most significant amplitudes in the sample. To make this statistic well defined it is necessary to regularize the peak counting. Here we adopt the simple prescription to demand peaks to be at least $\Delta \omega = 10$ apart from each other. To get the corresponding significances $\bar{A}_M$ as defined in Eq.~\eqref{eq:significance2} we again sample the PDF of $\hat{A}_{M}$ for each $M$ from the Fisher matrix.  

\begin{figure*}
\resizebox{0.8\hsize}{!}{
\includegraphics{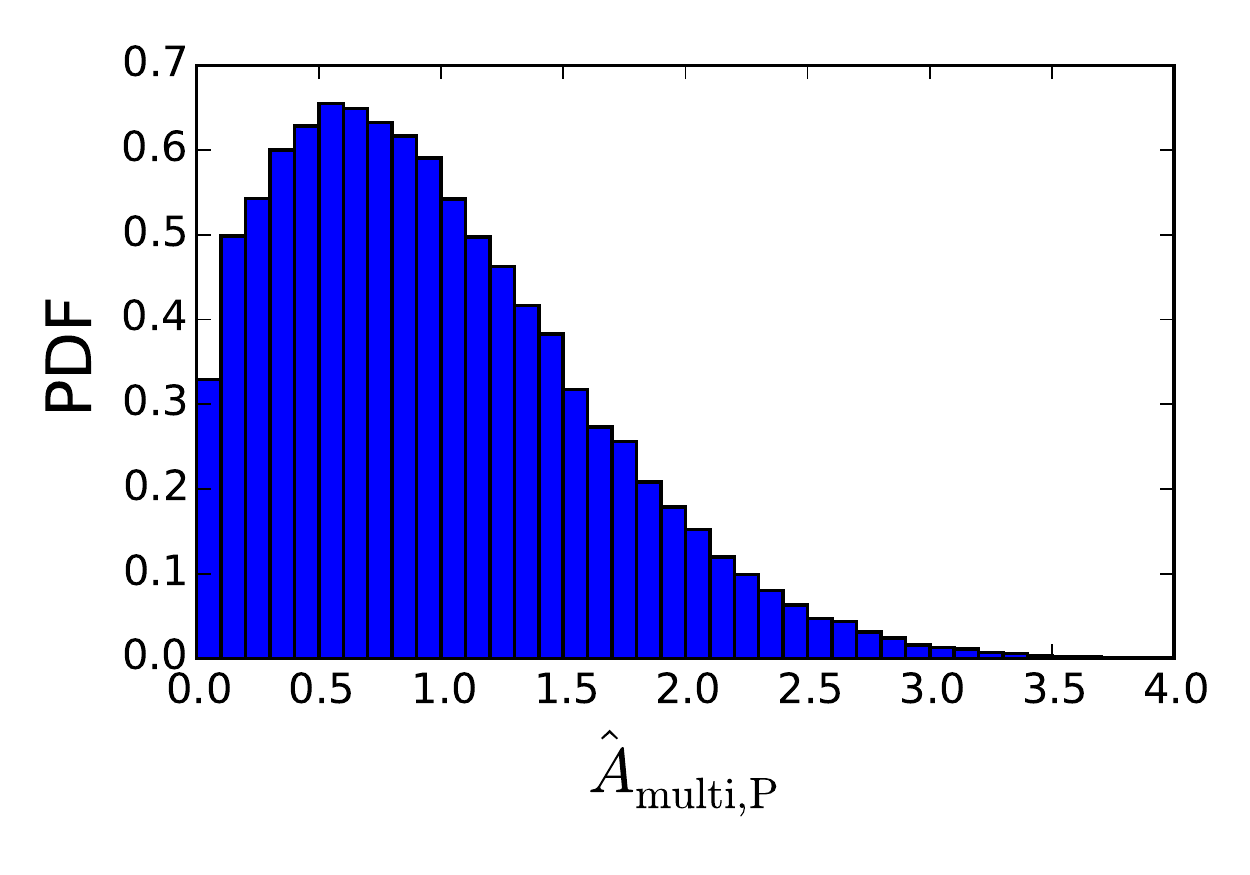}
\includegraphics{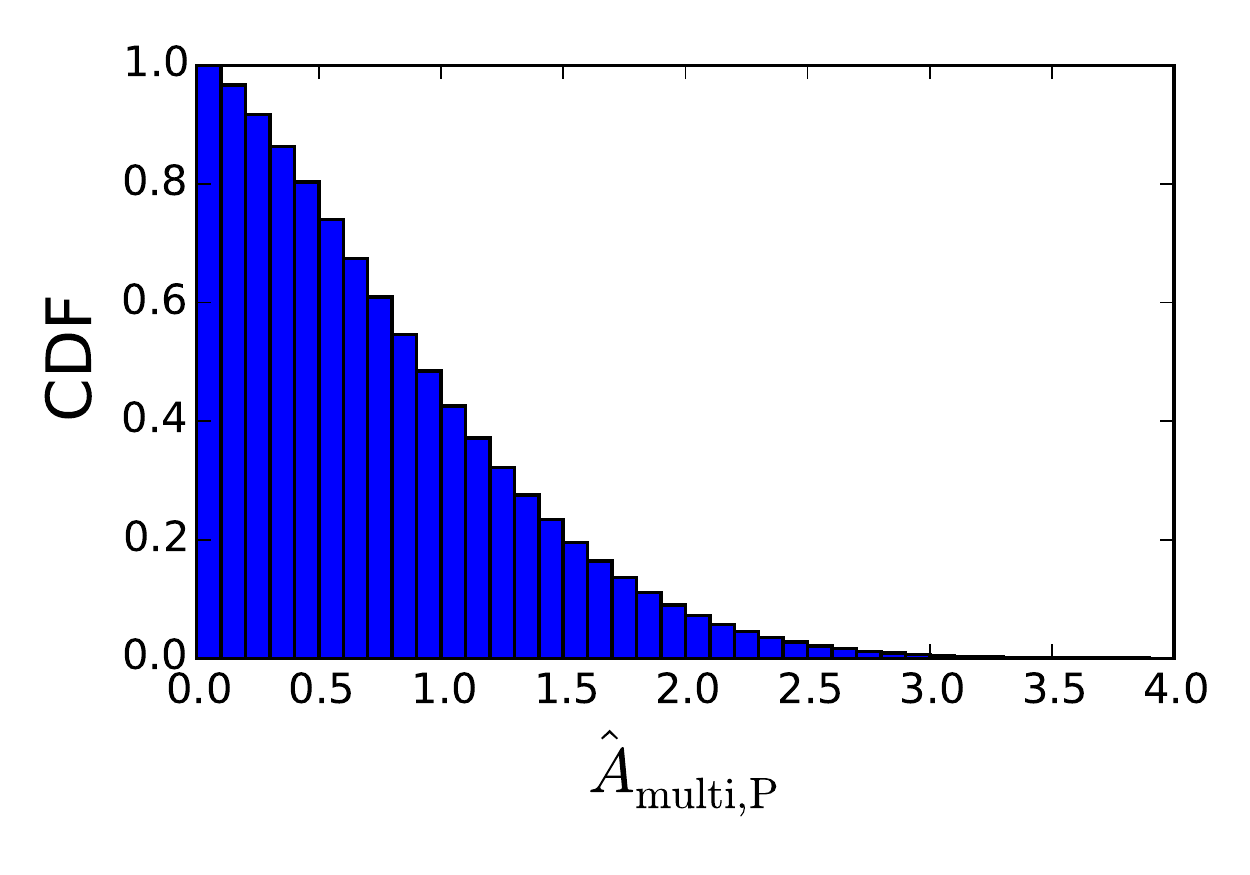}
}
\resizebox{0.8\hsize}{!}{
\includegraphics{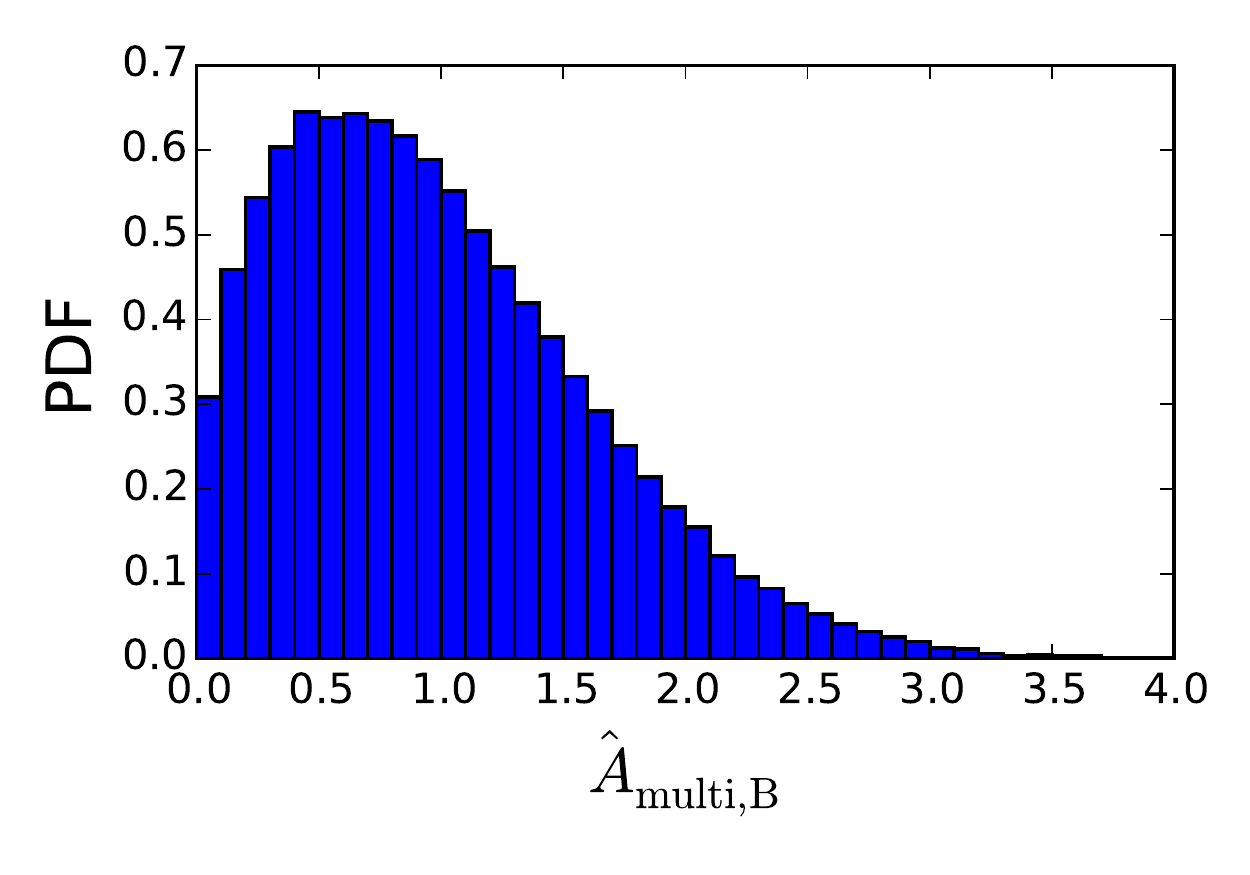}
\includegraphics{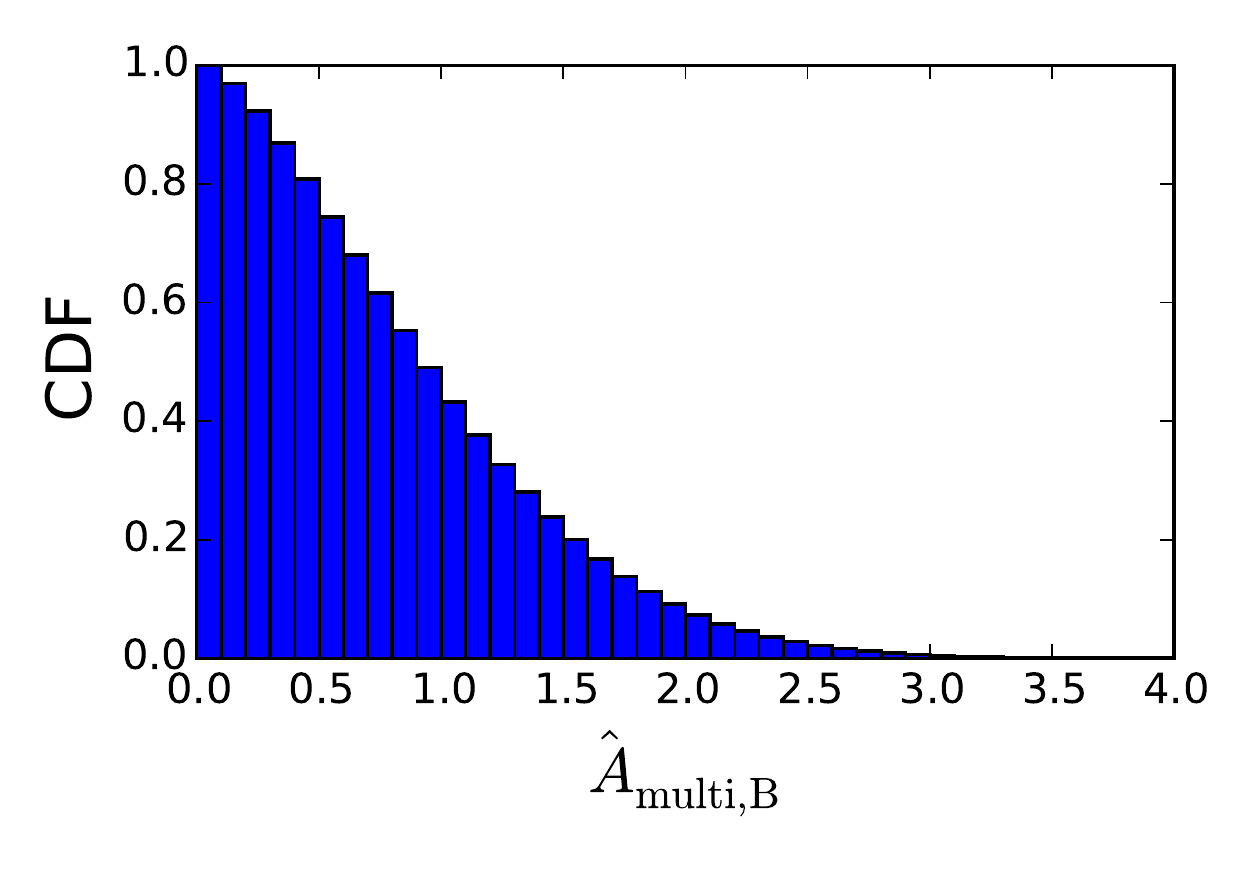}
}
\caption{Left: The PDF and CDF of the multi-frequency estimator eq.\eqref{eq:amaxmulti_single} for a power spectrum (top) and bispectrum (bottom) survey with with $0<\omega<1000$ and $M_{\text{max}}=10$.} 
\label{fig:multi_bispec}
\end{figure*}

Since the multi frequency models in \ref{sec:models} do not come with a prediction for $M$, in a second step we maximize over $M$ (with some predefined $M_{\text{max}}$), defining
\begin{equation}
\label{eq:amaxmulti_single}
\hat{A}_\mathrm{multi} = \max_M \bar{A}_M.
\end{equation}
As was pointed out in Ref.~\cite{ShellardBispectrumPsJoined2014} the maximization over peaks gives an additional look elsewhere effect. From the PDF for $\hat{A}_{\text{multi}}$ we get the multi peak significance $\bar{A}_M$ via eq.\eqref{eq:significance2}. An example with $M_{\text{max}}=10$ is shown in Fig.~\ref{fig:multi_bispec}. 

\section{Conclusions}
\label{sec:conclusion}

In this paper we have laid the foundation for a comprehensive combined search for resonance models in the power spectrum and bispectrum with Planck CMB data. Resonance models lead to unique challenges in CMB data analysis. This is in part because the highly oscillatory functions that appear have to be sampled very densely for an accurate numerical treatment. Further, the shape is non-separable and very difficult to expand with general modal techniques. For the power spectrum, the usual likelihood exploration is computationally challenging in the present case. We have shown that a simplified estimator based on a fixed $\Lambda$CDM cosmology is actually sufficiently close to the estimator that varies all cosmological parameters. We conclude that this is the result of a lack of significant correlations between the oscillatory parameters and those of the fiducial cosmology. 

We have developed a modal approach in the power spectrum, that can be used for all oscillating shapes, and will allow to cover a wide space of models very quickly. This method follows directly from our treatment of the bispectrum \cite{OptimalEstimator2014}, which allows for a fair comparison between the results of both analysis. 

The main point of this paper is the evaluation of the look-elsewhere effect in a combined search. While the necessary probability density functions have been previously characterized in Ref.~\cite{ShellardBispectrumPsJoined2014,Fergusson:2014tza}, in this paper we present a new method to sample these posterior probabilities. The results is sufficiently accurate and most importantly computationally feasible. We made use of the fact that the power spectrum and bispectrum shapes can be decomposed in their sine and cosine contributions, and that one can make a Gaussian approximation that describes these contributions for a given frequency sampling. This can be done for both power spectrum and bispectrum, and avoids running the estimator on a large amount of Monte Carlo maps, which would be almost impossible with our high frequency estimator. We use our sampling method to obtain some of the amplitude probability distribution functions for an idealized experiment with $\ell<2000$. With this study, we have all necessary tools for an application to the Planck data. Besides the resonance models which we focussed on in this paper, the methods presented here are also applicable to other types of rapidly varyings features in the power spectrum and bispectrum \cite{NGFeaturesChen2007,StepFeatureAdshead, Chen2015,CorrelatedFeaturesCs,CorrelatedFeaturesCs2,CorrelatedFeaturesCs3,CorrelatedFeaturesCs4,FeaturesFromHeavyPhysicsAna2011,NonBDBispectrum2009,NonBDBispectrum2010b,nonBDbispectrum2015} and should provide an efficient way to address multi-peak significance of these models.

\section*{Acknowledgments} 
We thank Fran\c{c}ois Bouchet for useful discussions and comments. PDM would like to thank the hospitality of the Alaska Sitka Sound Science Center were part of this work was completed. PDM would also like to thank Will Handley for the help with setting up Polychord. MM acknowledges funding by Centre National d'Etudes Spatiales (CNES). BDW is supported by a senior Excellence Chair by the Agence Nationale de Recherche (ANR-10-CEXC-004-01) and a Chaire Internationale at the Universit\'{e} Pierre et Marie Curie.

\vspace{1cm}

\appendix

\section{Scaling of the power spectrum}

\subsection{Large scales}

We asume 
\be
P = P_0(1+A \cos \omega \log k/k_*).
\ee
Therefore
\be
\delta \mathcal{C}_{\ell} \propto \int dk k^{-1} (k^{i \omega} + k^{-i \omega}) \Delta_{\ell}(k)^2,
\ee
with $\Delta_{\ell}(k)$ being the radiation transfer functions. The above integral can not be performed analytically, but on sufficiently large scales the transfer function is simply a geometrical factor $\Delta_{\ell}(k)\sim j_{\ell} (\Delta \eta k)$. 
Using this limit, we find 
\be
\delta \mathcal{C}_{\ell} \propto A (\Delta  \eta )^{-i \omega } \frac{\sqrt{\pi }  \Gamma \left(1-\frac{i \omega }{2}\right) 
   \Gamma \left(\ell +\frac{i \omega }{2}\right)}{4 \Gamma \left(\frac{3}{2}-\frac{i
   \omega }{2}\right) \Gamma \left(\ell -\frac{i \omega }{2}+2\right)} + {\rm c.c}.
\ee 
Besides the presence of oscillates, it is also clear that the projected amplitude $A_{\rm proj}$ is suppressed with respect to the primordial one. We can write \begin{widetext} 
\be
 \frac{\sqrt{\pi }  \Gamma \left(1-\frac{i \omega }{2}\right) 
   \Gamma \left(\ell +\frac{i \omega }{2}\right)}{4 \Gamma \left(\frac{3}{2}-\frac{i
   \omega }{2}\right) \Gamma \left(\ell -\frac{i \omega }{2}+2\right)} = \frac{ \Gamma \left(\frac{3}{2}+\frac{i
   \omega }{2}\right)}{ \Gamma \left(\frac{3}{2}-\frac{i
   \omega }{2}\right)} \frac{\sqrt{\pi }  \Gamma \left(1-\frac{i \omega }{2}\right) 
   \Gamma \left(\ell +\frac{i \omega }{2}\right)}{4 \Gamma \left(\frac{3}{2}+\frac{i
   \omega }{2}\right) \Gamma \left(\ell -\frac{i \omega }{2}+2\right)}.
\ee
We consider the limiting case $\omega \rightarrow \infty$ and find for the second term
\be
\lim_{\omega \rightarrow \infty} \frac{\sqrt{\pi }  \Gamma \left(1-\frac{i \omega }{2}\right) 
   \Gamma \left(\ell +\frac{i \omega }{2}\right)}{4 \Gamma \left(\frac{3}{2}-\frac{i
   \omega }{2}\right) \Gamma \left(\ell -\frac{i \omega }{2}+2\right)}  = \sqrt{\pi } (-1+i)\omega^{-5/2},
\ee
\end{widetext}
while the real part of the first term oscillates between $-1$ and $1$. Therefore, effectively $A_{\rm proj} \propto A \omega^{-5/2}$; the higher the frequency, the more the projected power spectrum will have its oscillations suppressed. 

\subsection{Small scales}

\begin{figure*}
\resizebox{0.6\hsize}{!}{
\includegraphics{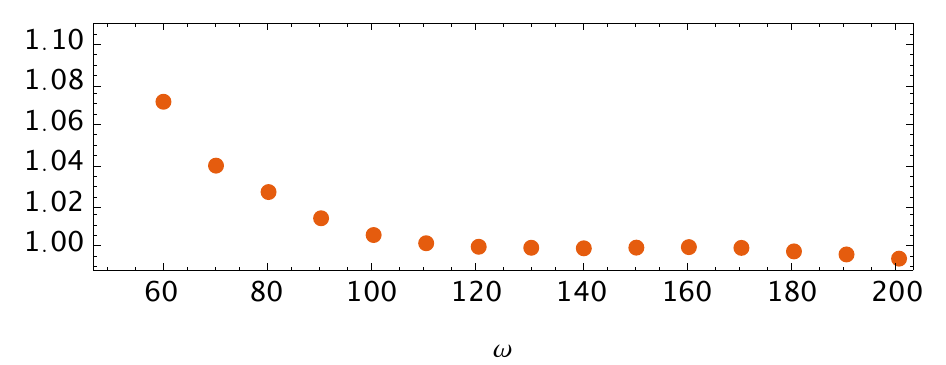}
}
\caption{Scaling of the projected power of the corrections to the power spectrum (resonance). The y-axis is a measure of the signal to noise of the correction (normalized to 1) multiplied by $\omega$.}
\label{fig:scalingAproj}
\end{figure*}

The computation above is only valid for very large scales. For small scales, we do not expect the scaling to hold. We can estimate the scaling by assuming that the transfer functions on small scales is simply a sine or cosine, suppressed by the damping factor. Since all modes are damped, we only care about it oscillating with some frequency set by the sound horizon at recombination \cite{SmallScales}. There is still the Bessel function, which has the limit $\lim_{\rightarrow \infty} j_{\ell}(x) = (x/\ell)^{\ell -1/2}/\ell$ (which turns out not to the affect the scaling).  We then have to perform integrals of the form
\be
\delta \mathcal{C}_{\ell} \propto \int dk k^{2(\ell -1)} (k^{i \omega} + k^{-i \omega}) \sin(r_s k)^2.
\ee
If we take $k_{\rm max} \rightarrow \infty$ the resulting integral diverges for $\omega \rightarrow \infty$. The reason is that in reality there is a damping scale, which we neglect. Instead we consider a hard cutoff $k_{\rm max}$. We obtain \begin{widetext}
\be
\frac{\mathit{k}_{\max }^{-i \omega +2 \ell -1} \left((-i \omega +2 \ell +1) \sin
   ^2\left(\mathit{k}_{\max } \mathit{r}_s\right)-2 \mathit{k}_{\max }^2 \mathit{r}_s^2
   \, _1F_2\left(\ell -\frac{i \omega }{2}+\frac{1}{2};\frac{3}{2},\ell -\frac{i \omega
   }{2}+\frac{3}{2};-\mathit{k}_{\max }^2 \mathit{r}_s^2\right)\right)}{-1+(2 \ell -i
   \omega )^2} + \rm{c.c}.
\ee
\end{widetext}
In the limit of large frequencies, the expression above decays as $1/\omega$. In other words, we expect that the amplitude of projected features is suppressed as $A_{\rm proj} \propto A \omega^{-1}$ on small scales. In Fig.~\ref{fig:scalingAproj} we show a measure of the total signal (to noise, with $\ell_{\rm min} = 200$) in $\delta \mathcal{C}_{\ell}$ multiplied by $\omega$, with some arbitrary normalization. Indeed we find that the signal roughly drops as $1/\omega$ with a steeper slope for lower frequencies.

Through numerical computations in Ref.~\cite{OptimalEstimator2014} we found that the projected amplitude of the bispectrum is similarly reduced as $f_{\rm NL}^{\rm proj} = f_{\rm NL} /\omega$. In EFT the expected amplitude of the bispectrum has an amplitude that grows as $\omega^{5/2}$ while the power spectrum only grows $\omega^{1/2}$. It is thus expected that for very large frequencies the bispectrum could become larger  primordially and in projection than the power spectrum. This is exactly where the EFT is expected to break down.

\bibliography{npointbib}

\end{document}